\documentclass[sigconf, nonacm]{acmart}

\usepackage{balance}
\usepackage{subcaption}
\usepackage{graphicx}
\usepackage{wrapfig}
\usepackage{amsmath}
\usepackage{paralist}
\usepackage{setspace}
\usepackage{comment}
\usepackage{xspace}
\usepackage{listings}
\usepackage[vlined,commentsnumbered,ruled,linesnumbered]{algorithm2e}
\usepackage{fixltx2e}
\usepackage{makecell}
\usepackage{mathtools} 
\usepackage{amsmath}
\usepackage{multirow}
\usepackage{float}
\usepackage{makecell}
\usepackage{url}
\usepackage{paralist}
\usepackage{tablefootnote}
\usepackage{hyperref}
\usepackage{tikz}
\usepackage{bbding}
\usepackage{makecell}
\usepackage{amsmath}
\usepackage[inline]{aplcomment}
\newcommenter{cong}{0.0,0.8,0.8}

\newtheorem{definition}{Definition}[section]

\newtheorem{lemma}{Lemma}[section]

\newtheorem{example}{Example}[section]

\newtheorem{query}{Query}[section]

\newcommand{\tool}{\textsc{PredTrace}\xspace}

\newcommand{\dbfield}[1]{\texttt{#1}}
\newcommand{\dbtable}[1]{\texttt{#1}}

\begin{document}
\title{Efficient Row-Level Lineage Leveraging Predicate Pushdown}
\author{Yin Lin
}
\affiliation{%
 \institution{
  University of Michigan\\
irenelin@umich.edu}
}
\authornote{Work done at Microsoft.}
\author{Cong Yan
}
\affiliation{%
 \institution{
  Microsoft Research\\
congyan.me@gmail.com}
}

\begin{abstract}
Row-level lineage explains what input rows produce an output row through a data processing pipeline, having many applications like data debugging, auditing, data integration, etc. 
Prior work on lineage falls in two lines: eager lineage tracking and lazy lineage inference. 
Eager tracking integrates lineage tracing tightly into the operator implementation, enabling efficient customized tracking. However, this approach is intrusive, system-specific, and lacks adaptability.
In contrast, lazy inference generates additional queries to compute lineage; it can be easily applied to any database, but the lineage query is usually slow. Furthermore, both approaches have limited coverage of the type of data processing pipeline supported due to operator-specific tracking or inference rules.
 
In this work, we propose \tool, a lineage inference approach that achieves easy adaptation, low runtime overhead, efficient lineage querying, and high pipeline coverage. 
It achieves this by leveraging predicate pushdown: pushing a row-selection predicate that describes the target output down to source tables and querying the lineage by running the pushed-down predicate. 
\tool may require saving intermediate results when running the pipeline in order to compute the precise lineage. When this is not viable, it can still infer lineage but may return a superset. Compared to prior work, \tool achieves higher coverage on TPC-H queries as well as 70 sampled real-world data processing pipelines in which UDFs are widely used. It can infer lineage in seconds, outperforming prior lazy approaches by up to 10x. 
\end{abstract}

\maketitle

\section{Introduction} \label{sec:intro}
Data lineage describes the origin from input data of an output data item produced by a data processing pipeline.
It has a number of applications in data debugging~\cite{gulzar2016bigdebug}, data integration~\cite{lenzerini2002data}, auditing~\cite{earley2015data}, GDPR compliance~\cite{truong2019gdpr}, interactive visualization~\cite{gershon1998information}, etc. For instance, a data scientist may wish to know how an erroneous output is produced and what potentially corrupted input may be responsible. A customer who is able to observe the output data may wish to delete her personal data from the entire pipeline, relying on the system's support of tracking a particular output through the pipeline till the data source. These operations are expressed as lineage queries over the workflow, to determine the subset of input records that produce some given output.

Finding out lineage for SQL queries has been widely explored by prior work, which mainly divides into two approaches: {\it eager lineage tracking} which proactively tracks data dependency during query execution~\cite{psallidas2018smoke,interlandi2015titian, widom2004trio, bhagwat2005annotation, karvounarakis2013collaborative}, and {\it lazy lineage inference} that reasons about lineage and constructs queries to retrieve related data from the input~\cite{glavic2009perm,cui2000tracing,cui2003lineage, glavic2009provenance, ikeda2013logical}.
Eager approaches add lineage tracking overhead during query runtime and require changing the data platform to support the tracking. 
Prior work explores various techniques to make the tracking efficient~\cite{psallidas2018smoke, gulzar2016bigdebug, interlandi2015titian}, but each is deeply integrated with a particular database system~\cite{psallidas2018smoke}. Such deep integration makes it hard to apply to other platforms, achieve the same performance, or evolve together with the constantly changing DBMS itself. In contrast, lazy inference generates a new query to obtain both the query result and its lineage without interfering with the original query execution. However, each lineage query incurs a significant computation overhead as it often recomputes the original query with additional lineage computation, which can be hundreds of times slower than the original query itself~\cite{glavic2009perm, glavic2009provenance}.

Furthermore, both eager and lazy approaches often focus on simple queries with a few relational operators, such as SPJA queries. This is because eager tracking tailors its tracking approach for each type of operator, while lazy inference utilizes operator-specific rules to construct the lineage query. As a result, both approaches face challenges in effectively supporting complex queries, let alone non-relational operators and user-defined functions (UDF) commonly found in data processing pipelines~\cite{predicate-pushdown}. 
Such a limitation greatly restricts the lineage application to simple SQL queries but renders it impossible or impractical for complicated SQL queries or data processing pipelines.

In this work, we propose \tool, an efficient system to infer row-level lineage. It requires little intervention with the data processing systems while facilitating efficient lineage querying. The key insight is to leverage predicate pushdown, a well-studied query optimization technique. Briefly, to logically infer the lineage of a particular output row $t_o$, \tool uses a {\it row-selection predicate} in the form of $col_1=v_1\land col_2=v_2\land ...$ to represent $t_o$, where $v_i$ representing the corresponding column values, and pushes it down to the data pipeline's source tables to select the lineage rows.

When a predicate cannot be pushed down at an operator or does not select the precise lineage, 
\tool resolves by saving an intermediate result. The intermediate result enables replacing the failed predicate with a row-selection predicate, which is guaranteed to succeed by \tool. This adds little overhead to the pipeline runtime and can be easily implemented in most systems. \tool further optimizes the intermediate result by reducing its size. When the user queries the lineage for a specific output, \tool runs the corresponding predicate on the intermediate result and the source tables, which is usually very efficient. Furthermore, since predicate pushdown can be applied not only to complex relational queries~\cite{hellerstein1993predicate, levy1997query, zhou2021sia} but also to data pipelines with non-relational operators and user-defined functions (UDFs)~\cite{predicate-pushdown}, it enables \tool to infer lineage for such queries and pipelines, providing much wider coverage compared to prior work.

We recognize that saving intermediate results may not always be feasible. To tackle this challenge, we extend \tool to {\it compute lineage by directly running predicates on the source tables}, without relying on any intermediate results.
We believe that this solution is practical under real-world constraints. For instance, lineage queries may need to be performed on a finished pipeline that cannot be easily reproduced, or on a continuously running pipeline that spans across multiple engines and cannot be interrupted.

In summary, this work makes the following contributions.
\begin{asparaitem}
    \item We formally discuss how predicate pushdown can assist lineage inference. Specifically, we demonstrate that pushing down a row-selection predicate through a single operator yields a predicate that precisely selects the lineage, and pushing it down through a pipeline assisted by saving intermediate results always computes the precise lineage.  
    \item We introduce optimizations to reduce the size of intermediate results, resulting in substantial storage savings of up to 99\% and a significant reduction in lineage query time by up to 99\%.
    \item We present an alternative lineage solution when intermediate result materialization is not feasible. 
    This solution is practical as it only requires query logs and infers lineage by executing predicates on source tables. It may return a lineage superset, and we propose an improved algorithm to reduce the false positives. Our evaluation demonstrated its effectiveness across all TPC-H queries and pipelines in our benchmark, with an average FPR as low as 6.6\%.
    \item We build a prototype system \tool and evaluate it on TPC-H queries and 70 real-world data pipelines sampled from open-source Jupyter Notebooks. \tool computes lineage for all TPC-H queries and 70 pipelines, outperforming previous work \cite{cui2003lineage, ikeda2013logical, niu2017provenance, glavic2009provenance} that handle only a subset and exhibit low efficiency. 
    If intermediate results can be saved, \tool is $98 \times$ more efficient in lineage querying than the best prior lazy lineage system. If intermediate results cannot be saved, it is still $6 \times$ more efficient. 
\end{asparaitem}
\vspace{-2mm}
\section{Background: Predicate Pushdown}\label{sec:background}

Predicate pushdown is a widely adopted query optimization technique used to reduce the data loaded before executing a query. Pushing a predicate through an operator $Op$ can be formalized as follows: given a predicate $F$ on $Op$'s output, we want to find the most selective predicate $G$ on $Op$'s input, such that $Op(G(T))=F(Op(T))$ holds for any table $T$. With this, pushing a predicate through a query can be treated as pushing it operator-by-operator following the data flow represented by its logical plan.

Existing query optimizers can already identify predicate pushdown opportunities for relational queries~\cite{hellerstein1993predicate, zhou2021sia, levy1997query}. Recent work like MagicPush~\cite{predicate-pushdown} extends this to data science pipelines with non-relational operators and UDFs. MagicPush employs a search-verification approach: it searches for the most selective $G$ (e.g., by combining all predicates in $F$ and $Op$), then verifies if $Op(G(T)) = F(Op(T))$. It returns the first candidate that passes the verification. MagicPush can be configured to relax the criteria when equivalence cannot be achieved, returning the first $G$ that satisfies $F(Op(G(T))) = F(Op(T))$, where $G$ helps reduce the input table but $F$ is still needed. MagicPush designed a new verification mechanism that works on a bounded-size symbolic table (e.g., a table of two rows) using symbolic execution but can guarantee correctness on tables of any size. Therefore, it can handle a large set of UDFs as long as they can be symbolically executed.

In this work, we utilize an external predicate pushdown module, such as MagicPush, which returns a $G$ satisfying either $Op(G(T)) = F(Op(T))$ or $Op(G(T)) \supseteq F(Op(T))$. However, leveraging predicate pushdown alone cannot solve the lineage tracing problem, as it does not provide precise lineage information and may result in large supersets of the actual lineage. To address this, we establish a connection between predicate pushdown and lineage tracing, demonstrating that precise lineage can be obtained when intermediate results are materialized.  Furthermore, when materialization is not feasible, our approach significantly reduces false positives.

\section{Proposed Approach: \tool}\label{sec:approach}

\subsection{Running example}
We demonstrate \tool using the following example of TPC-H Q4,  as demonstrated below. The query output of Q4 is shown in Figure \ref{fig:q4} (2).

\vspace{1.5mm}
\noindent
\texttt{SELECT }\textit{o\_orderpriority, count(*) as order\_count} \\
\hspace*{1em}\texttt{FROM} \textit{orders} \\
\hspace*{0.5em}\texttt{WHERE} \textit{o\_orderdate > `1993-07-01'} \texttt{AND} \textit{o\_orderdate < `1993-10-01'}\\
\hspace*{1.5em}\texttt{AND} \texttt{exists(}\\
\hspace*{3em}\texttt{SELECT *} \\
\hspace*{4em}\texttt{FROM} \textit{lineitem}\\
\hspace*{3.5em}\texttt{WHERE} \textit{l\_orderkey = o\_orderkey} \\
\hspace*{4.5em}\texttt{AND} \textit{l\_commitdate < l\_receiptdate})\\
\hspace*{1em}\texttt{GROUP BY} \textit{o\_orderpriority} \\
\hspace*{1em}\texttt{ORDER BY} \textit{o\_orderpriority}; 

\vspace{-1.5mm}
\begin{example} [Row-level data lineage]
    After executing Q4, a user may want to trace the input rows that contributed to the generation of a specific output row, such as $t_o$: <\texttt{o\_orderpriority} = '1-URGENT', \texttt{order\_count} = 10594>. By applying the algorithms presented later, the user can identify the lineage of $t_o$, including rows from the \texttt{orders} table that contribute to the '1-URGENT' group, as well as the rows from the \texttt{lineitem} table involved in the nested query that influences the presence of the \texttt{orders} lineage rows.
\end{example}

\subsection{Foundations of lineage computation} \label{sec:foundations}
We extend the lineage model from prior work~\cite{ikeda2013logical} to compute lineage in pipelines containing non-relational operators and UDFs.

\begin{figure}[]
    \centering
    \includegraphics[scale=0.39]{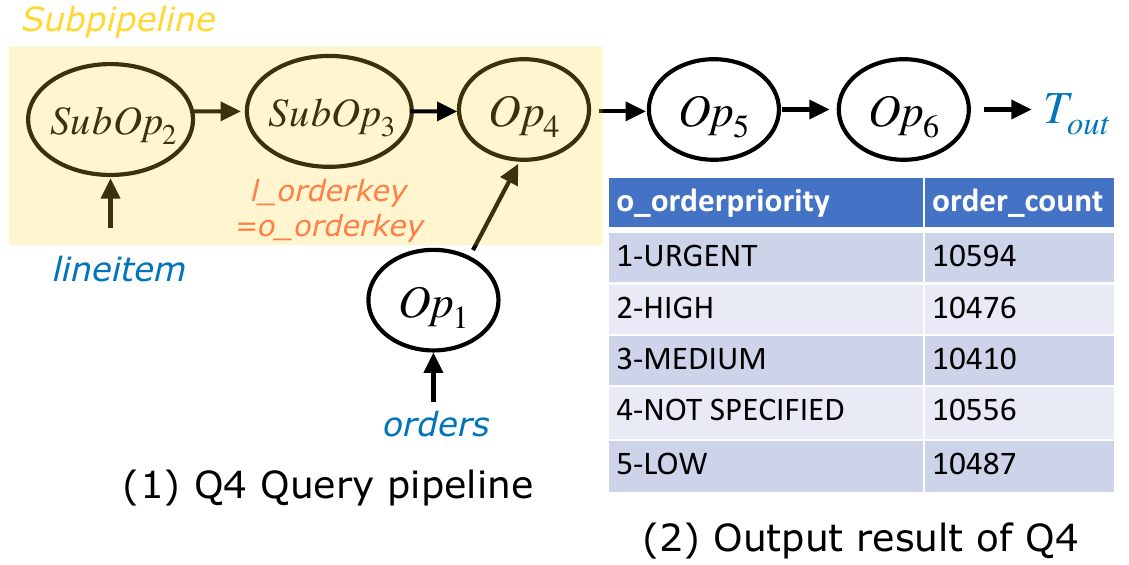}
    \vspace{-2mm}
    \caption{\textnormal{Query pipeline and output result for TPC-H Q4.}}
    \label{fig:q4}
    \vspace{-4mm}
\end{figure}

\noindent
\textbf{Data pipelines.} 
A data pipeline consists of sequential processing steps, each transforming input tables $T$ into an output table $T_{out}$ for analytical queries or data preparation. These steps are represented by \textit{operators} connected through a dataflow. We focus on Pandas \cite{pandas} operators and operators supported by DBMSs such as PostgreSQL, as summarized and listed in Table \ref{tab:basic-ops}.

\begin{example}
    TPC-H Q4 can be expressed as a data pipeline in \tool's syntax, as shown in Figure \ref{fig:q4} (1). The pipeline consists of the following transformations: $SubOp_{2,3}$ denotes the Filter operators applied to the \dbtable{lineitem} table within the subquery (\dbfield{l\_commitdate}<... and \dbfield{l\_orderkey}=...). $Op_4$ is a SemiJoin operator performing a semijoin with the subquery.  $Op_{1,5,6}$ represent the Filter (\dbfield{o\_orderdate}>...), GroupBy, and Sort operators applied to the main query on the \dbtable{orders} table. Operator descriptions are listed in Table \ref{tab:q4}.
\end{example}

\noindent
\textbf{Lineage model.}
We first define data lineage for a single operator. 

\begin{definition} [Single operator lineage] \label{def:precise_lineage}
    Let $Op$ be an operator over tables $T$ which produces an output table $T_{out}$ (i.e., $Op(T)=T_{out}$). For a record $t_o \in T_{out}$, its lineage is the minimal subset of $T$ (denoted as $T^*$) such that $Op(T^*) \supseteq \{t_o\}$. In other words, $t_o$'s lineage is a minimal subset of $Op$'s inputs that can produce $t_o$. If the minimal subset is not unique, the lineage is the union of all minimal subsets.
\end{definition}

The $\supseteq$ symbol is used in the definition because, in 1-to-N or N-to-N transformations (e.g., \textit{Unpivot}, \textit{RowExpand}, and \textit{WindowOp} in Table \ref{tab:basic-ops}), no single subset of the input can exclusively produce the output $t_o$.  For instance, with \textit{RowExpand}, each input row expands into multiple output rows, making it impossible for a single input row to solely produce $t_o$.

We next discuss the lineage model for a data pipeline.

\begin{definition} [Lineage of a data pipeline]
    Consider a data pipeline consisting of operators \( Op_1, Op_2, \ldots, Op_n \). The dataflow of the pipeline is represented as \( (Op_1 \circ Op_2 \circ \cdots \circ Op_n)(T_0) = T_n \), where \( T_0 \) is the pipeline input and \( T_n \) is the output table. For a record $t_o$ in $T_n$, the lineage of $t_o$ is the minimal subset of $T_0$ (denoted as $T_0^*$) such that \( (Op_1 \circ Op_2 \circ \cdots \circ Op_n)(T_0^*) \supseteq t_o \). 
\end{definition}

\noindent
\textbf{Lineage inference.}
We begin by illustrating how predicate pushdown infers lineage for a single operator. We define a \textit{row-selection predicate}, denoted as $F^{row}$, which is constructed from a concrete row $t$ as $(col_1 == v_1 \wedge col_2 == v_2 \wedge \cdots)$, where $v_1, v_2, \cdots$ are the corresponding values from $t$. In other words, $F^{row}(Op(T)) = \{t\}$ under set semantics. Given an output record $t_o$, we construct $F^{row}$ and push it through $Op$.

Assuming that every intermediate result is materialized, obtaining pipeline lineage via predicate pushdown is straightforward. We push down $F^{row}_i$ through $Op_i$ to obtain $G^{row}_i$, apply $G^{row}_i$ on the input of $Op_i$, and convert the resulting rows into $F^{row}_{i-1}$. This process continues until we reach the source tables. Next, we prove the correctness of this approach.

\begin{lemma}
    The pushed-down predicate $G^{row}_1$ selects the precise lineage from the input $T_0$. Specifically, (1) it identifies the minimal subset of $T_0$ that (2) can produce the output $t_o$.
\end{lemma}
\vspace{-3mm}
\begin{proof} [Proof]
We first address property (2). For a single operator, if the $G^{row}$ pushed down from $F^{row}$ is the most selective predicate\footnote{There is no need to consult a cardinality estimator to decide selectivity, as $G^{row}$ is a conjunction of equal comparisons; therefore, the predicate with the most conjunctions is the most selective.} that ensures $Op(G^{row}(T)) \supseteq F^{row}(Op(T))$, then $G^{row}$ selects the precise lineage as defined in Definition \ref{def:precise_lineage}. 
In the context of a pipeline, the situation becomes more complex, as the lineage model is not transitive. Specifically, if $Op_{i}(T_{i-1}^*)$ returns more rows than the target output, it produces a superset of $T_{i}^*$. In such cases, when the subsequent operator $Op_{i+1}$ is a {\it non-monotonic} operator \cite{monotonic-query}, applying $Op_{i+1}$ to a superset of $T_{i}^*$ does not guarantee the production of the expected target output $T_{i+1}^*$. However, by employing predicate pushdown, the row-selection predicate $F_i^{row}$ at each operator's output ensures that only the target output is produced, such that $F_i^{row}(Op_{i}(T_{i-1}^*))=T_{i}^*$ for any table $T_i$.  This means that $F_i^{row}$ can effectively reduce any superset $Op_{i}(T_{i-1}^*)$  to the target $T_{i}^*$.

For property (1), since the $F^{row}$ at each operator selects the exact target output and $G^{row}$ is the most selective pushed-down predicate, $G^{row}$ will therefore ensure the selection of the minimal lineage set. 
\end{proof}

However, materializing every intermediate result is prohibitively expensive. Instead, we can skip saving these results by pushing down $F_n^{row}$ from $T_n$ until we reach the source tables.

\vspace{-1mm}
\begin{lemma} \label{lemma2}
    Pushing down $F_n^{row}$ to the source tables without using intermediate results yields a $G_1$ that (1) selects a superset of the precise lineage from $T_0$ and (2) can produce the target output $t_o$.
\end{lemma}
\vspace{-3mm}
\begin{proof} [Proof]
We begin by demonstrating property (2), which is crucial for ensuring the lineage superset is meaningful, as opposed to being a random superset.  Formally, we aim to show that $Op_i(G_i(T_i)) \supseteq T_{i+1}^*$, where $T_{i+1}^*$ is the precise lineage from $T_{i+1}$, the target output at operator $Op_i$. We prove this by induction.
Assuming that $G_{i+1}(T_{i+1})\supseteq T_{i+1}^*$ holds, property (2) is satisfied because the predicate pushdown mechanism ensures that $Op_i(G_i(T_i))\supseteq F_i(Op_i(T_{i}))$. Since $F_i$ is the predicate pushed down from operator $Op_{i+1}$, i.e., $F_i = G_{i+1}$, we have $Op_i(G_i(T_i)) \supseteq G_{i+1}(T_{i+1}) \supseteq T_{i+1}^*$.

Given that $G_{i+1}(T_{i+1})\supseteq T_{i+1}^*$ holds by assumption, property (1) is also trivially satisfied. Since $T_i^*$ is the minimum subset that can produce the target output $T_{i+1}^*$, any other subset of $T_i$ that produces $T_{i+1}^*$ must necessarily be a superset of $T_i^*$. 
\end{proof}

Directly pushing down $F_n^{row}$ can yield a meaningful lineage superset, but this superset may become impractically large. In \tool, we propose saving intermediate results only when pushing down $G_{i+1}$ is not equivalent to pushing down $F^{row}_i$, ensuring precise lineage tracing. If intermediate results cannot be saved, \tool incorporates a procedure that significantly reduces false positives while still providing a meaningful superset.

\subsection{\tool workflow}

\begin{algorithm}[t]
	\DontPrintSemicolon
	\SetKwInOut{Input}{input}\SetKwInOut{Output}{output}
   \SetKwFunction{GetRowSelection}{\textsc{GetRowSelection}}
   \SetKwFunction{Concretize}{\textsc{Concretize}}
	\LinesNumbered
	\Input{\textit{data pipeline $Op_1, Op_2, \cdots, Op_n$, output record $t_o$}}
	\Output{\textit{ $T_i^*$ for all source tables $T_i$} } \BlankLine
    \tcc{Logical lineage inference phase}
    $F_n^{row} \leftarrow$ parameterized row-selection predicate \\
 \ForEach{operator $Op_i$ in reverse topological order}{
         $G_i \leftarrow$  Push down $F_i$ through $Op_i$ \\
        \If{$G_i$ does not select precise lineage}{
        $F^{row}_{i}\leftarrow$ \GetRowSelection($Op_i$) \\ 
        {$G_i^{row} \leftarrow$ Push down $F_i^{row}$ through $Op_i$
        }}}
    \tcc{Pipeline execution phase}
    \If{no intermediate result is required}{Run the pipeline without modification}\Else{Run the modified pipeline to save intermediate results}
 \tcc{Lineage querying phase}
        \ForEach{pushed-down predicate $G^{T_i}$ on source table $T_i$}
        {$G^{T_i} \leftarrow$ \Concretize($t_o$)\\
        ${T_i^*} \leftarrow$ Run the concretized $G^{T_i}$ on $T_i$}
	\Return \textit{$T_i^*$ for all $T_i$}
	\caption{Row-level lineage tracing in \tool}
 \label{alg:overview}
\end{algorithm}

We illustrate the workflow of \tool in Algorithm \ref{alg:overview}.
Given a data pipeline, the process begins with the \textit{logical lineage inference phase}. Initially, a parameterized row-selection predicate, $F_n^{row} = (col_1 == v_1 \wedge col_2 == v_2 \wedge \cdots)$, is constructed, where each ${col}_i$ represents an output column and $v_i$ is the corresponding value (Line 2). \tool then pushes $F_n^{row}$ down through the operators using a predicate pushdown module, as described in Section \ref{sec:background}. At each operator $Op_i$, it first attempts to push $F_i$ (i.e., $G{i+1}$), which may not be a row-selection predicate, through $Op_i$ (Line 4). \tool then relies on a verification process to determine whether pushing down $F_i$ is equivalent to pushing down a row-selection predicate. If the verification process determines that the resulting $G_i$ does not select precise lineage, the output of $Op_i$ needs to be materialized, and \tool pushes down a row-selection predicate $F^{row}_i$ instead (Lines 5-7).

We note that the logical lineage inference operates independently of the underlying data system, as it relies solely on the pushdown of a parameterized row-selection predicate (with the $v_i$s as variables). This allows lineage inference to be performed once per query or pipeline, regardless of subsequent lineage queries.

In the \textit{pipeline execution phase}, if no intermediate results are required, \tool runs the pipeline as is (Lines 9-10). Otherwise, it modifies the pipeline to save intermediate results (Lines 11-12). For a DBMS executing a relational query, \tool splits the query into multiple queries. For platforms running data pipelines written in an imperative language like Python, \tool rewrites the code to save the intermediate results.

During the \textit{lineage querying phase}, the user specifies an output row $t_o$ for lineage tracing. \tool uses $t_o$ to concretize the pushed-down predicates (Line 15).
If no intermediate results are saved, the pushed-down predicates consist only of variables from $t_o$. If an intermediate result is saved at $Op_i$, the pushed-down predicates include variables from $F^{row}_i$. \tool then executes the original $F_i$ on the saved results, replacing variables in $F^{row}_i$ with the corresponding rows. This approach is similarly applied when multiple intermediate results are saved. Finally, \tool runs each $G^{T_i}$ on source table $T_i$ to obtain the lineage rows (Line 16).

The remainder of this paper is structured as follows: Section \ref{sec:lineage-inference} introduces the pushdown of row-selection predicates and the verification process for determining equivalent pushdowns. Section \ref{sec:opt} discusses techniques for optimizing the size of intermediate results. Finally, Section \ref{sec:minimize_superset} presents an algorithm to minimize lineage supersets when materializing intermediate results is not feasible.

\vspace{-2mm}
\subsection{Example workflow: TPC-H Q4}
We explain the workflow using TPC-H Q4. Continuing with the running example, and given the pipeline of Q4, \tool begins with the \textit{logical lineage inference phase}. It constructs a parameterized row-selection predicate, $F_6^{row} = (\dbfield{o\_orderpriority} == v_g \wedge \dbfield{order\_count} == v_s)$, which is to be pushed down through the Sort operator ($Op_6$). The pushed-down predicate $G_6$ is the same as $F_6^{row}$ and selects the precise lineage, therefore \tool continues. It performs similarly for the next GroupBy operator ($Op_5$), obtaining $G_5$ as \dbfield{o\_orderpriority}==$v_g$.

\begin{table}[tp]
\caption{\textnormal{Operators and predicate descriptions for TPC-H Q4.}}\label{tab:q4}
\vspace{-3mm}
\centering
\small
\begin{tabular}{|l|l|l|}
\hline
\textbf{Op} & \textbf{Descr.} & \textbf{Predicate}\\
\hline
$Op_6$& Sort & \makecell[l]{$F_6^{row}$ :o\_orderpriority==$v_g \wedge$ \\{} {} {} {} {} {} {} {} {} {} {} {} {} {} 
 order\_count==$v_s$} \\
\hline 
$Op_5$& GroupBy & \makecell[l]{$F_5$ / $G_6$: o\_orderpriority==$v_g \wedge$\\ {} {} {} {} {} {} {} {} {} {} {} {} {} {} order\_count==$v_s$} \\
\hline
$Op_4$&  \makecell[l]{SemiJoin} & \makecell[l]{$F_4$ / $G_5$: 
 o\_orderpriority==$v_g$ \\$F_4^{row}$:o\_orderkey==$v_o\wedge$\\{} {} {} {} {} {} {} {} {} {} o\_orderpriority==$v_g$}\\
\hline
$SubOp_3$& \makecell[l]{Filter\\ (l\_orderkey=...)} & \makecell[l]{$F_3$ / $G_4$: o\_orderkey==$v_o$} \\
\hline
$SubOp_2$& \makecell[l]{Filter\\ (l\_commitdate<...)} &  \makecell[l]{$F_2$ / $G_3$: l\_orderkey==$v_o$}  \\
\hline
\dbtable{lineitem}& Source table & \makecell[l]{$G_2^l$: l\_orderkey==$v_o\wedge$\\ {} {} {} {} {} {} {} {} l\_commitdate<...} \\
\hline
$Op_1$& \makecell[l]{Filter\\ (o\_orderdate>...)} & \makecell[l]{$F_1$ / $G_4$: o\_orderkey==$v_o$} \\
\hline
\dbtable{orders}& Source table & \makecell[l]{$G_1^o$: o\_orderkey==$v_o\wedge$\\{} {} {} {} {}{} {} {} o\_orderdate>...} \\
\hline
\end{tabular}
\vspace{-5mm}
\end{table}

At the next SemiJoin operator, $Op_4$, the predicate $F_4$ involves only the \dbfield{o\_orderpriority} column. As a result, pushing $F_4$ down produces a \texttt{True} predicate on \dbtable{lineitem}, which is not equivalent to pushing down a row-selection predicate that selects the precise lineage. If a row-selection predicate were pushed down, the join column \dbfield{o\_orderkey} would be included to filter the correlated rows from the \dbtable{lineitem} table. 
Therefore, \tool determines that an intermediate result of $Op_4$ needs to be saved and pushes down a row-selection predicate $F^{row}_4$ instead. Since $F^{row}_4$ contains \dbfield{o\_orderkey}, pushing it down to the subquery ($SubOp_{3}$ and $SubOp_{2}$) returns $G_2^l$=(\dbfield{l\_orderkey}==$v_o\wedge$\dbfield{l\_commitdate}<...) on the \dbtable{lineitem} table, which selects the precise lineage. Pushing $F^{row}_4$ down to the main query yields $G_4$=\dbfield{o\_orderkey}==$v_o$ on the output of $Op_1$. This is subsequently pushed down as $F_1$, which produces the final $G^o_1=$( \dbfield{o\_orderkey}==$v_o\wedge$\dbfield{o\_orderdate}>...) on the \dbtable{orders} table.

In the \textit{pipeline execution phase}, since \tool has decided that the intermediate result of $Op_4$ needs to be saved, it splits the original query into two parts to materialize this result, as shown below and then runs the rewritten queries.

\noindent
Q4-1: \texttt{SELECT }\textit{o\_orderkey, o\_orderpriority} \texttt{INTO} \textbf{\textit{inter\_table}}\\
\hspace*{3em}\texttt{FROM} \textit{orders} \\
\hspace*{2.5em}\texttt{WHERE} \textit{o\_orderdate>=\'1993-07-01\'} \texttt{AND} \textit{o\_orderdate<\'1993-10-01\'}\\
\hspace*{3.5em}\texttt{AND} \texttt{exists(}\\
\hspace*{5em}\texttt{SELECT *} \texttt{FROM} \textit{lineitem}\\
\hspace*{5.5em}\texttt{WHERE} \textit{l\_orderkey = o\_orderkey} \\
\hspace*{6.5em}\texttt{AND} \textit{l\_commitdate < l\_receiptdate}); \\
Q4-2: \texttt{SELECT }\textit{o\_orderpriority, count(*) as order\_count} \\
\hspace*{3em}\texttt{FROM} \textbf{\textit{inter\_table}} \\
\hspace*{2.5em}\texttt{GROUP BY} \textit{o\_orderpriority}\\
\hspace*{2.5em}\texttt{ORDER BY} \textit{o\_orderpriority}; 
\noindent
\vspace{1mm}

After the query completes, in the \textit{lineage querying phase}, if the user queries the lineage of the output row $F_6^{row}$= (\dbfield{o\_orderpriority}\\=='1-URGENT' and \dbfield{order\_count}==10594), \tool concretizes and runs the original $F_4$ on the intermediate result of $Op_4$, returning rows with \dbfield{o\_orderkey} in (193, 1350, ...). Then it replaces $v_o$ in $G^l_2$ and $G^o_1$ with these orderkeys, and runs them on the corresponding source tables to obtain the lineage rows.

\section{Lineage inference leveraging predicate pushdown}
\label{sec:lineage-inference}

\subsection{Row-selection predicate pushdown}\label{sec:pushdown}

\begin{table*}
  \caption{\textnormal{Core operators of \tool, what UDF is supported, and the default $G^{row}$ used if pushing down $F^{row}$ fails.}}
\scriptsize{
\vspace{-0.15in}
  \label{tab:basic-ops}
  \begin{centering}
  \scalebox{1.17}{
  \setlength{\tabcolsep}{1.5pt} 
  \begin{tabular}{lllll}
  \toprule
  {\bf Operator} & {\bf Description} & {\bf embedded UDF supported} & {\bf default $G^{row}$ if pushdown fails} & {\bf default $G^{row}$ explained} \\
  \midrule
  Filter & selection in SQL & UD-selection predicate & $F^{row}\wedge$ filter-predicate & Select the same row as $t_o$\\
  \hline
  InnerJoin & innerjoin in SQL & UD-join predicate & \makecell[l]{$col_{l_1} == v_{l_1} \wedge \cdots$ (left)\\ $col_{r_1} == v_{r_1} \wedge \cdots$ (right)} & \makecell[l]{Select the rows from left/right table which\\produce $t_o$ after joined} \\
  \hline
  RowTransform &  row/scalar transform & UD-transform function & $f_1(col_1,col_2\cdots) == v_1\wedge\cdots$ & \makecell[l]{Select the row that transformed\\into the values of $t_o$}  \\
  \hline
  DropColumn &  projection in SQL  & NA & $F^{row}$ & Select the same row as $t_o$   \\
  \hline
  Reorder/TopK & order-by (LIMIT N) in SQL & UD-compare function  & $F^{row}$ & Select the same row as $t_o$  \\
  \hline
  Union/Intersect & union/intersection in SQL & NA & $F^{row}$ & Select the same row as $t_o$ \\
  \hline
  GroupBy & groupby in SQL & UD-aggregation & $col_{group} == v_i$ & Select the entire group\\
  \hline
  Pivot &  pivot operator & UD-aggregation & $col_{index} == v_i\wedge$ & Select the entire index group  \\
  \hline
  UnPivot  & unpivot operator & NA &  $col_{index}==v_1\wedge col_{v_2}==v_3$ & Select the entire index group  \\
  \hline
  RowExpand &  1-to-k transformation & UD-expand function & \makecell[l]{$(f_1(col_1,col_2,\cdots)[0] == v_1\vee$\\$f_1(col_1,col_2,\cdots)[1] == v_1) \wedge \cdots$} & \makecell[l]{Select the row if any row from its\\expanded K-row output is the same as $t_o$} \\
  \hline
  LeftOuterJoin & left/right outer join & UD-join predicate &\makecell[l]{$col_{l_1} == v_{l_1} \wedge \cdots$ (left)\\ $col_{r_1} == v_{r_1} \wedge \cdots$ or $False$ (right)} & \makecell[l]{Select the row matching $t_o$ from left table,\\the joined row from right table if \\right column values from $t_o$ are not NULL,\\o.w. select empty set from right table}\\
  \hline
  WindowOp & rolling/diff like ops & NA &  $col_{index}\in [i,i+$window\_size$]$ & Select the entire window used to produce $t_o$\\
\midrule
  GroupedMap & \makecell[l]{transform grouped sub-\\tables with a subquery} &  \makecell[l]{Subquery defined in\\ core operators} & $col_{group} == v_i$ & Select the entire group \\
    \hline
  SemiJoin & EXIST/IN subquery & \makecell[l]{Subquery defined using\\core operators} & \makecell[l]{$F^{row}$ (outer)\\push $True$ to inner query, (inner)\\replace correlated column $col_i$ with $v_i$} & \makecell[l]{Select same $t_o$ from outer table;\\select all rows returned by inner query with\\specific correlated column value $v_i$ from $t_o$} \\
  \hline
  AntiJoin & NOT EXIST subquery &  \makecell[l]{Subquery defined using\\core operators} & \makecell[l]{$F^{row}$ (outer)\\ push $False$ to inner query} & \makecell[l]{Select same $t_o$ from outer table\\empty set from inner query result}  \\
    \hline
  SubQuery & \makecell[l]{For each row, the subquery\\computes with other tables\\to return a value $v$} & \makecell[l]{Subquery defined using\\core operators} & \makecell[l]{$F^{row}$ (outer)\\push $col_v==v$ to inner query, (inner)\\replace correlated column $col_i$ with $v_i$} & \makecell[l]{Select same $t_o$ from outer table\\all rows used to produce value $v$ from inner table} \\
  \bottomrule
\end{tabular}}
\end{centering}
}
 \vspace{-0.1in}
\end{table*}

As demonstrated in Section \ref{sec:foundations}, identifying the most selective $G^{row}$ when pushing down a row-selection predicate $F^{row}$ is crucial for precise lineage selection.
\tool relies on an external pushdown module like MagicPush \cite{predicate-pushdown}, which typically returns the most selective $G^{row}$ when pushing down $F^{row}$ through $Op$. To handle cases where the external module might fail, \tool provides default $G^{row}$ values as listed and explained in Table~\ref{tab:basic-ops}.  Usually, the default $G^{row}$ is the same as the $G^{row}$ obtained through predicate pushdown.
However, there are scenarios where the default $G^{row}$ may differ, returning a lineage superset for certain UDFs.
For example, in a groupby operation with max aggregation, the precise lineage includes only the row with the maximum value, rather than the entire group.

In our experiment, MagicPush consistently pushed down $F^{row}$ successfully, and the corresponding $G^{row}$ is the most selective predicate returning the precise lineage. Therefore, \tool rarely needs to fall back on the default $G^{row}$ in practice.

\vspace{-1mm}
\subsection{Verification of equivalent pushdown}
\label{sec:verification-nonrowsel}
\label{sec:smt-minimal}
\begin{figure}
    \centering
    \includegraphics[scale=0.42]{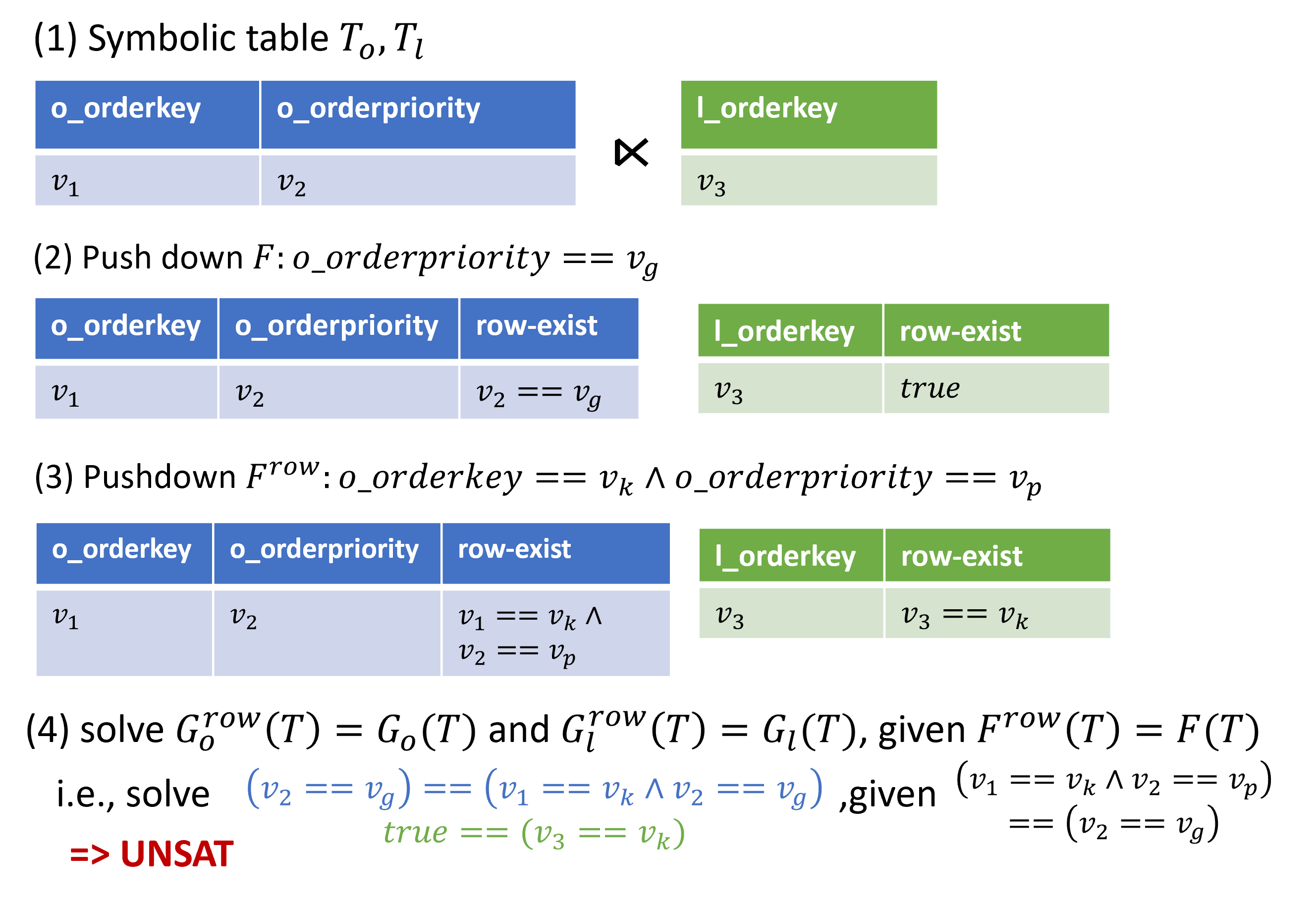}
    \vspace{-9mm}
    \caption{\textnormal{Verification for non-row-selection predicate.}}
    \vspace{-5mm}
    \label{fig:counter-ex-minimal}
\end{figure}

Next, we illustrate how \tool determines whether pushing down a non-row-selection predicate yields an equivalent result to pushing down a row-selection predicate. Briefly, given an arbitrary predicate $F$, and its corresponding pushed-down predicate $G$, we can construct a row-selection predicate $F^{row}$ from $F$, push it down as $G^{row}$, and verify if $G$ is equivalent to $G^{row}$.

This verification process begins with the construction of symbolic input tables of fixed size, where cell values are represented by symbols such as $v_1, v_2, v_3$. For instance, consider the SemiJoin ($Op_4$) in Q4, where \tool attempts to push down a non-row-selection predicate $F =$ \dbfield{o\_orderpriority} == $v_g$. We first create symbolic tables, each containing a single tuple with cell values represented as $v_1, v_2, v_3$, as shown in Figure \ref{fig:counter-ex-minimal}(1).

To verify equivalence, \tool first pushes down the non-row-selection predicate $F$ to obtain $G$, applying $G$ symbolically to the corresponding tables. This process uses \textit{row-exist} expressions to capture the conditions required for rows to pass the filter. These expressions track whether each row satisfies the filtering conditions set by $G$. For the predicate $F = $ \dbfield{o\_orderpriority} == $v_g$, pushing it down results in $G_o = $ \dbfield{o\_orderpriority} == $v_g$ on the \dbtable{orders} table and $G_l = $ \texttt{True} on the \dbtable{lineitem} table. Applying these predicates to the symbolic tables updates the \textit{row-exist} symbol for each row, as shown in Figure \ref{fig:counter-ex-minimal}(2).

\tool then constructs a $F^{row}$, pushes it down to obtain $G^{row}_o$ and $G^{row}_l$, and similarly applies them to their corresponding tables, with the results shown in Figure \ref{fig:counter-ex-minimal}(3).

Finally, \tool verifies whether $G$ is equivalent to $G^{row}$ when $F^{row}$ is equivalent to $F$. This involves checking whether the corresponding result tables, particularly the {\it row-exist} expressions, are equivalent regardless of the values of $v_{1,2,3}$. In the example, the verification indicates that $G_l$ is not equivalent to $G^{row}_l$ (i.e., \texttt{True} is not equivalent to $v_3==v_k$). Therefore, we cannot use $G$ and must fall back to saving the output of $Op_4$ and push down $F^{row}$.

Predicate equivalence is undecidable in first-order logic, but the SMT solver returns sound results when it finds equivalence. While it may time out, our evaluation showed it consistently produced sound results in practice.

\subsection{Scope and limitations}

\textbf{Set semantics.} Since \tool relies on the row-selection predicate to identify the target output, it tracks all rows with the same value indistinguishably, returning their lineage together. 
While this may limit applications like view updates, where users need to delete a specific output, \tool remains useful in scenarios like debugging, where the focus is on outlier values rather than individual rows.
In our evaluation, since each table/dataframe has a unique index by default, they inherently follow set semantics. 
For a given pipeline, a primary key could be added to make each output row distinguishable at every step, which we leave for future work.

\textbf{Scope of UDF supported.} \tool can support UDF embedded in an operator (e.g., user-defined aggregation in groupby) as described in Table~\ref{tab:basic-ops}, but not arbitrary user-defined operators. This limitation is primarily due to the constraints of the predicate pushdown module~\cite{predicate-pushdown}, which requires the UDF to be 1) deterministic, 2) symbolically executable \cite{veanes2009symbolic}, and 3) cannot allow library functions when embedded in certain operators. 
Since \tool heavily depends on the SMT solver, complex UDF verification may lead to significant overhead or even timeouts. However, in our evaluation, all verifications were completed within a reasonable time.

\section{Intermediate result optimizations}
\label{sec:opt}

In this section, we illustrate two optimizations to reduce the size of the materialized intermediate results. Smaller intermediate results reduce storage and processing overhead and speed up running $F_i$ to concretize $F^{row}_i$ during lineage querying. The first optimization locally projects only the necessary columns for intermediate results. The second optimization globally searches for potential intermediate results at later operators with smaller sizes.

\textbf{Column projection. } Not all columns are required in the materialized intermediate result. \tool retains only two types of columns that contain essential information for running the pipeline and inferring precise lineage. The first type includes columns used in later operators. For instance, in the intermediate result of $Op_4$ in Q4, the \dbfield{o\_orderpriority} column is used in the GroupBy ($Op_5$) and Sort ($Op_6$) operations. The second type includes columns necessary for equivalently pushing down $F^{row}_i$, such as primary keys, join keys, and correlated columns in subqueries. As a result, column projection in Q4's intermediate result retains only two columns: \dbfield{o\_orderpriority} and \dbfield{o\_orderkey}.

\tool then rewrites the row-selection predicate $F^{row}_i$ to involve only the projected columns. If the rewritten predicate fails to push down, unlike the original predicate, it reverts back to the original predicate, but we do not see this happen in practice.

\begin{figure}[tp]
    \centering
    \includegraphics[scale=0.35]{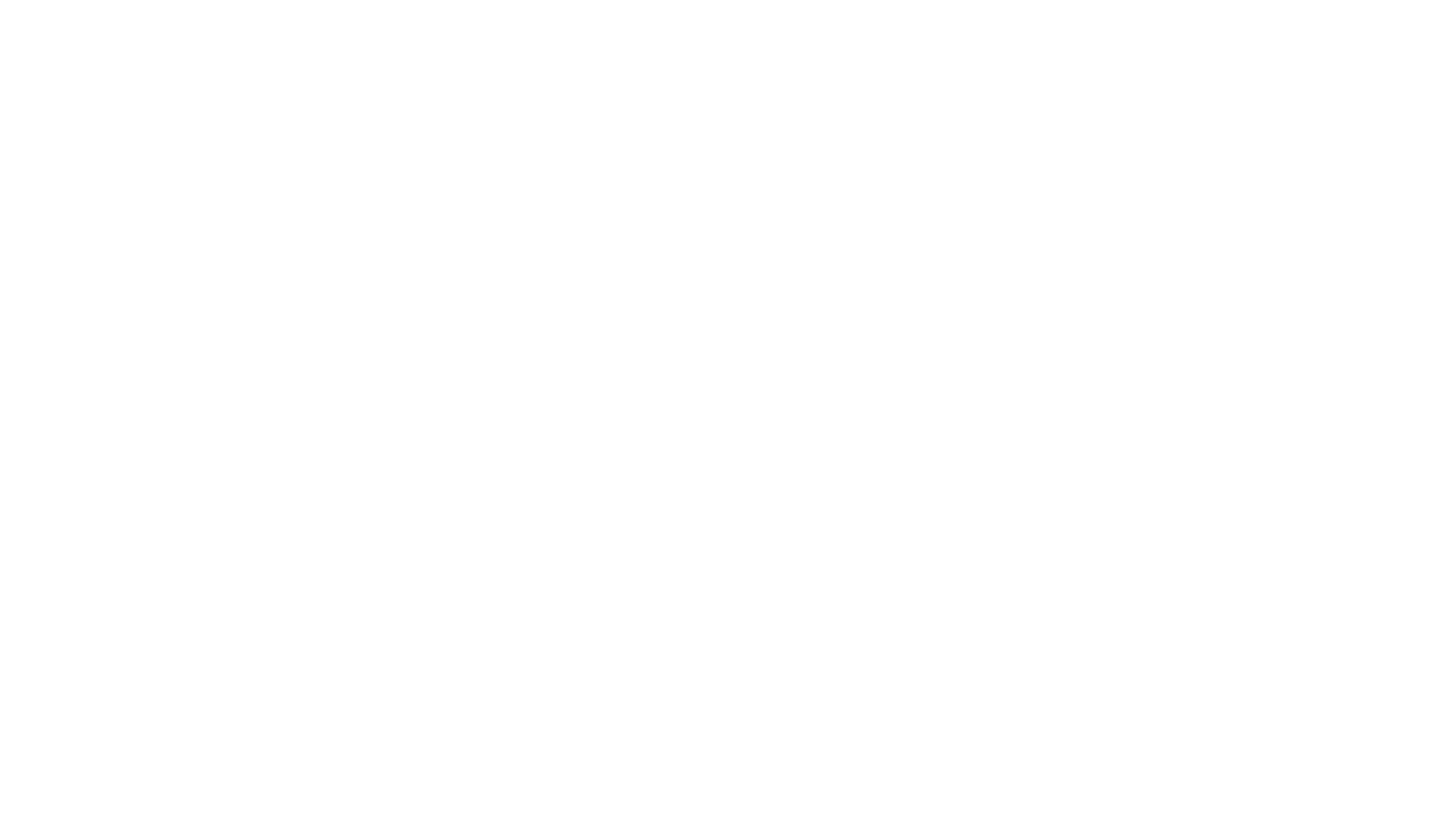}
    \vspace{-4mm}
    \caption{\textnormal{Data pipeline of TPC-H Q3.}}
    \label{fig:q3}
    \vspace{-4mm}
\end{figure}

\textbf{Choosing intermediate results. } In certain cases, deferring the materialization of intermediate results until a later operator in the pipeline can also result in successful pushdowns at all earlier operators. To make this choice, we identify all alternative intermediate results and then leverage size estimation from the DBMS to choose the intermediate result with the minimal size.

We illustrate this using TPC-H Q3. Figure \ref{fig:q3} shows the pipeline of Q3. A row-selection predicate, expressed as \texttt{l\_orderkey} = $v_1 \wedge$ \texttt{o\_orderdate} = $v_2 \wedge$ \texttt{o\_shippriority} = $v_3$, can be pushed down directly through $Op_4$ without needing to save an intermediate result. However, when attempting to push it through the next InnerJoin operator ($Op_3$), \tool determines that an intermediate result after $Op_3$ should be saved because the join column \texttt{c\_custkey} is missing in $F_3$, which would result in the entire \dbtable{customer} table being included in the lineage. To optimize the process, an alternative is to save the intermediate result after $Op_4$, which achieves the same effect. By turning $F_4$ into a row-selection predicate $F^{row}_4$, $F_3$ would then include \texttt{c\_custkey}, and pushing it down would return the precise lineage. Additionally, the output of $Op_4$ has a significantly smaller join size compared to that of $Op_3$.

Algorithm \ref{alg:opt} outlines the optimization process. It starts by performing column projection on the intermediate result of $Op_i$ (Lines 1-2). Then it explores alternative intermediate results after $Op_i$ (Line 3). For each candidate $T_j$, the algorithm assumes materializing it, projects the necessary columns of $T_j$ (Lines 4-5), and obtains a row-selection predicate $F^{row}_{j}$(Line 6). If $F^{row}_{j}$ can be pushed down to obtain the same lineage as saving the naive intermediate result (i.e., pushing down $F^{row}_{i}$), and the column-projected $T_j$ is smaller, it is considered a viable alternative (Lines 7-8). Otherwise, the algorithm stops and returns the optimal intermediate result found so far (Lines 9-11).

The algorithm has complexity $O(n)$ for a data pipeline of length $n$. In our approach, we estimate the intermediate result size based on the physical plan generated by the DBMS. When applicable, materialization is often deferred until after filters or joins, as these operations significantly reduce the data size by filtering or joining relevant rows. This approach resulted in a storage reduction of over 99\% in our evaluation.

\begin{algorithm}[h]t]
	\DontPrintSemicolon
	\SetKwInOut{Input}{input}\SetKwInOut{Output}{output}
 	\SetKwFunction{ColumnProjection}{\textsc{ColumnProjection}}
   \SetKwFunction{ObtainInterResult}{\textsc{ObtainInterResult}}
   \SetKwFunction{MaterializeIntermediateResult}{\textsc{MaterializeIntermediateResult}}
   \SetKwFunction{ObtainRowSelection}{\textsc{ObtainRowSelection}}
   \SetKwFunction{CanPushDown}{\textsc{CanPushDown}}
	\LinesNumbered
	\Input{ \textit{naive\_inter\_table} on the output of operator $Op_i$}
	\Output{ \textit{optimized\_inter\_table}} \BlankLine
    $ColP_i \leftarrow$ \ColumnProjection(\textit{naive\_inter\_table}) \\
    \textit{optimized\_inter\_table} $\leftarrow \alpha_{ColP_i}$(\textit{naive\_inter\_table}) \\
 \ForEach{operator $Op_j$ after $Op_i$}
        {$T_j \leftarrow$ \MaterializeIntermediateResult($Op_j$) \\
        $ColP_j \leftarrow$ \ColumnProjection($T_j$) \\
        $F^{row}_{j}\leftarrow$ \ObtainRowSelection($\alpha_{ColP_j}(T_j)$) \\
        \If{\CanPushDown($F^{row}_{j}$) and est\_size($\alpha_{ColP_j}(T_j)$)  < est\_size(\textit{optimized\_inter\_table})}
        {\textit{optimized\_inter\_table} $\leftarrow \alpha_{ColP_j}(T_j)$}
        \Else{break}
        }
	\Return \textit{optimized\_inter\_table}
	\caption{Intermediate result optimization}
 \label{alg:opt}
\end{algorithm}

\section{Without saving intermediate result} \label{sec:minimize_superset}

Naively pushing down predicates can result in supersets that are too large to be useful (e.g., including the entire \dbtable{lineitem} table in Q4). In this section, we present an algorithm that utilizes additional predicate pushup and pushdown phases to reduce the superset size, leading to zero false positives for most pipelines.

Intuitively, when pushing down a predicate at $Op_i$ selects a lineage superset, we continue the pushdown process.  Once it reaches the source tables, we create a predicate representing the lineage rows for each table and push it up. During this pushup process, the lineage information from the source tables, such as the column values for \text{o\_orderkey} on the \dbfield{orders} table, is used to derive more selective predicates at join-like operators (e.g., SemiJoin $Op_4$). By pushing these refined predicates down again, we can produce more specific predicates, such as (\dbfield{l\_orderkey} IN $\Pi_{\text{o\_orderkey}}(\dbfield{orders})$) on the \dbfield{lineitem} table, filtering out non-joinable false positives.

\vspace{-1mm}
\subsection{Predicate pushup} \label{sec:pushup}

We first define the notion of a \textit{row-value predicate} that is pushed up from the source tables. When the pushdown process reaches $G^{T_i}$ at source table $T_i$, \tool generates a parameterized row-value predicate for each input table, denoted as $G^{T_i}\!\uparrow$. Specifically, $G^{T_i}\!\uparrow$ is expressed as $G^{T_i}\!\uparrow = col^{T_i}_1 \in \mathbb{V}^{T_i}_1 \wedge \cdots$, where $\mathbb{V}^{T_i}_1$ represents a set of values for column $col^{T_i}_1$, and this predicate is subsequently pushed up.
The predicate pushup process "simulates" computing the pipeline using only the lineage rows returned by $G^{T_i}$. Therefore, when pushing up predicates $G_i\!\uparrow,\cdots$ through $Op_i$, we want to find the $F_i\!\uparrow$ that precisely exhibits the output computed from input $G_i\!\uparrow(T_i)$, i.e., the $F_i\!\uparrow$ that satisfies Eqn~\eqref{eqn:pushup}:

\vspace{-1mm}
\begin{footnotesize}
\begin{equation} \label{eqn:pushup}
Op_i\left(G_i\!\uparrow\left(T_i\right)\right) = F_i\!\uparrow\left(Op_i\left(G_i\!\uparrow\left(T_i\right)\right)\right)
\vspace{-1mm}
\end{equation}
\end{footnotesize}

We use a search-verification approach to find $F_i\!\uparrow$: we enumerate candidates for $F_i\!\uparrow$, and for each candidate, we verify whether Eqn \eqref{eqn:pushup} holds, returning the first candidate passing the verification. 

To generate candidates, we rewrite each $G_i\!\uparrow$ into a predicate that can be applied to $Op_i$'s output:
\begin{asparaitem}
\item For any column that is removed from $Op$'s output, like a projection operator that retains only a subset of the columns, we remove all comparisons involving that column from $G_i\!\uparrow$.
\item If $Op$ computes value for a new column, like projecting \dbfield{a}+\dbfield{b} as column \dbfield{c}, we extract all equal comparisons from $G_i\!\uparrow$ that involve columns used for computation (e.g., \dbfield{a}$\in \mathbb{V}_1$ and \dbfield{b}$\in \mathbb{V}_2$). If all such columns match a specific value in $G_i\!\uparrow$, we generate a new comparison involving only the new column.  
\end{asparaitem}
After the above transformations, we generate the first candidate by conjuncting all the transformed $G_i\!\uparrow$s, then enumerate the remaining candidates by removing one comparison from the first candidate each time. In practice, the first candidate often passes the verification and we do not need to check other candidates. 

We verify Eqn \eqref{eqn:pushup} using the same process as described in Section~\ref{sec:smt-minimal}. 
First, we construct symbolic tables for the input $T_i$ and compute $Op_i(G_i\!\uparrow(T_i))$ symbolically. Then, we check whether the result is equivalent when applying $F_i\!\uparrow$ to the symbolic output using an SMT solver.

\subsection{Algorithm}

\begin{algorithm}[t]
	\DontPrintSemicolon
	\SetKwInOut{Input}{input}\SetKwInOut{Output}{output}
 	\SetKwFunction{ColumnProjection}{\textsc{ColumnProjection}}
   \SetKwFunction{IsSuperset}{\textsc{IsSuperset}}
   \SetKwFunction{IsMonotonic}{\textsc{IsMonotonic}}
   \SetKwFunction{Filter}{\textsc{Filter}}
	\LinesNumbered

    \tcc{Phase 1: pushdown} 
    $F_n^{row} \leftarrow$ parameterized row-selection predicate \\
    \ForEach{operator $Op_i$ in reverse topological order}
        {
            {$G_i\leftarrow$ Push down $F_i$ allowing a superset, ensuring $F_i\left(Op_i\left(G_i(T_i)\right)\right) = F_i\left(Op_i(T_i)\right)$}\\
        }
    \tcc{Phase 2: pushup} 
    \ForEach{source table $T_i$} 
        {
                $G^{T_i}\!\uparrow\leftarrow (col^{T_i}_1\in \mathbb{V}^{T_i}_1\wedge col^{T_i}_2\in \mathbb{V}^{T_i}_2\wedge\cdots)$
        }
    \ForEach{operator $Op_i$ in topological order}
        {
        $F_i\uparrow\leftarrow$ Push up $G_{i^1}\!\uparrow\cdots$ through $Op_i$ \\
        }
    \tcc{Phase 3: pushdown again} 
    \ForEach{operator $Op_i$ in reverse topological order}
        {
            $ F_i\!\downarrow \leftarrow F_i\!\downarrow \wedge F_i \wedge F_i\!\uparrow $ \\
            $G_i\!\downarrow \leftarrow$ Push down $F_i\!\downarrow$ through $Op_i$\\
        }
    \tcc{Phase 4: Iterative refinement} 
    {\it VValue} $\leftarrow$ empty map\\
    \ForEach{source table $T_i$}
    {
        ${T_i^{*\supseteq}}\leftarrow$ Run the concretized $G^{T_i}$ on $T_i$ \\
        \ForEach{$\mathbb{V}^{T_i}_1, \mathbb{V}^{T_i}_2\cdots$ in $G^{T_i}\!\uparrow$}
        {
            {\it VValue}[$\mathbb{V}^{T_i}_k$]$\leftarrow$ Project $col^{T_i}_k$ on $T_i^{*\supseteq}$
        }
    }
    \While {any $T_i^{*\supseteq}$ is updated during the last iteration}
    {   
        \ForEach{source table $T_i$}
        {
            Replace $\mathbb{V}^{T_j}_k, \cdots$ in $G^{T_i}\!\downarrow$ with {\it VValue}[$\mathbb{V}^{T_j}_k$] $, \cdots$ \\
            ${T_i^{*\supseteq}}\leftarrow$ Run the concretized $G^{T_i}\!\downarrow$ on $T_i$ \\
            {\it VValue}[$\mathbb{V}^{T_i}_k$]$\leftarrow$ values of column $col^{T_i}_k$ in $T_i^{*\supseteq}$
        }
    }
    \Return{$T_i^{*\supseteq}$ for all $T_i$}
	\caption{Iterative lineage refinement}
 \label{alg:opt-no-snapshot}
\end{algorithm} 


The algorithm is outlined in Algorithm~\ref{alg:opt-no-snapshot}, which consists of four phases. It begins with phase 1, pushing down predicates. We are not restricted to finding a $G_i$ that returns the precise lineage (i.e., equivalent to $G_i^{row}$). 
We use MagicPush to obtain a pushdown predicate that allows a lineage superset, i.e., $G_i(T_i) = T_i^{*\supseteq}$ (where $T_i \supseteq T_i^{*\supseteq} \supseteq T_i^*$), ensuring $F_i\left(Op_i\left(G_i(T_i)\right)\right) = F_i\left(Op_i(T_i)\right)$ (Line 4). This results in a meaningful superset, as shown in Lemma \ref{lemma2}.

In phase 2, for each source table, the algorithm initializes a parameterized row-value predicate (Lines 6-7), where $\mathbb{V}^{T_i}$ represents a set of values. At each operator $Op_i$, we apply the procedure outlined in Section \ref{sec:pushup} to derive a pushed-up predicate $F_i\!\uparrow$ (Lines 8-9).

In phase 3, the algorithm pushes down $F_n^{row}$ again. At each operator, $F_i \!\downarrow$ is conjuncted with $F_i$ from phase 1 and the pushup predicate $F_i \!\uparrow$ from phase 2. This more selective predicate is then pushed down, allowing only equivalent pushdown (Lines 11-13).

While the previous phases operate on parameterized predicates, phase 4 concretizes them by first running $G^{T_i}$ on the source tables (Line 17) to map variables like $\mathbb{V}^{T_i}_k$ to concrete column values (Lines 18-19). This mapping is stored in {\it VValue}. Since the final pushed-down predicate $G^{T_i}\!\downarrow$ may involve variables from other tables, the concretized variables further refine the lineage set selected by $G^{T_i}\!\downarrow$, reducing the value sets of its own variables (Lines 22-24). The algorithm runs iteratively until the value sets of all variables no longer shrink (Line 20). The remaining rows are returned as a superset or equivalent of the precise lineage.

The time complexity of lineage inference (phases 1-3) is $O(n)$, where $n$ is the length of the pipeline. This process is executed once per query or pipeline. During lineage querying, \tool only runs phase 4 for different target output rows, without repeating the earlier phases. The time complexity of lineage querying (phase 4) depends on the number of iterations required for all value sets to converge to a fixpoint. This is influenced by the pipeline's join structure, with the number of iterations bounded by the length of the longest sequence of joins. 
Therefore, the time complexity for iterative lineage querying is $O(l \times d)$, where $l$ is the length of the longest join sequence and $d$ is the size of the input data.

\begin{table}[tp]
\caption{\textnormal{Predicate descriptions when no intermediate result is used.}}\label{tab:q4-no-snapshot}
\vspace{-4mm}
\centering
\small
\scalebox{0.93}{
\begin{tabular}{|l|l|l|}
\hline
\textbf{Op} & \textbf{Predicate in phase 1: push down}\\
\hline
$SubOp_3$& $G_4$ / $F_3$: True \\
\hline
$Op_1$& \makecell[l]{$G_4$ / $F_1$: o\_orderpriority==$v_g$} \\
\hline
\dbtable{orders}& $G_1^o$: o\_orderpriority==$v_g\wedge$ o\_orderdate>... \\
\hline
\dbtable{lineitem}& $G_2^l$: l\_commitdate<l\_receiptdate \\
\hline
\hline
\textbf{Op} & \textbf{Predicate in phase 2: push up}\\
\hline
\dbtable{orders} & $G_1^o\!\uparrow$: o\_orderkey$\in \mathbb{V}^o_{\text{ o\_orderkey}}\wedge\cdots$\\
\hline
\dbtable{lineitem} & \makecell[l]{$G_2^l\!\uparrow$: l\_orderkey$\in \mathbb{V}^l_{\text{l\_orderkey}}\wedge\cdots$} \\
\hline
$Op_4$&  \makecell[l]{$G_{4^1}\!\uparrow$/$F_1\!\uparrow$: o\_orderkey$\in \mathbb{V}^o_{\text{ o\_orderkey}}\wedge\cdots$\\$G_{4^2}\!\uparrow$/$F_3\!\uparrow:$ l\_orderkey$\in \mathbb{V}^l_{\text{l\_orderkey}}\wedge\cdots$} \\
\hline
$Op_5$& \makecell[l]{$G_5\!\uparrow$/$F_4\!\uparrow$: (o\_orderkey$\in \mathbb{V}^o_{\text{ o\_orderkey}}\wedge\cdots$)$\wedge$\\{} {} {} {} {} {} {} {} {}{}{} {}{}{}(o\_orderkey$\in \mathbb{V}^l_{\text{l\_orderkey}}$)} \\
\hline
\hline
\textbf{Op} & \textbf{Predicate in phase 3: push down again}\\
\hline
\dbtable{orders}& \makecell[l]{$G_1^o\!\downarrow$: (o\_orderkey$\in \mathbb{V}^l_{\text{l\_orderkey}}$)$\wedge$\\{} {} {} {}
((o\_orderkey$\in \mathbb{V}^o_{\text{ o\_orderkey}}\wedge\cdots$)$\wedge$(o\_orderpriority==$v_g$...)) } \\
\hline
\dbtable{lineitem}& \makecell[l]{$G_2^l\!\downarrow$: (l\_orderkey$\in \mathbb{V}^o_{\text{ o\_orderkey}}$)$\wedge$\\{} {} {} {}
((l\_orderkey$\in\mathbb{V}^l_{\text{l\_orderkey}}\wedge\cdots$)$\wedge$(l\_commitdate<...))} \\
\hline 
\hline
\textbf{Iteration} & \textbf{Values of set variable in phase 4: iterative refinement} \\
\hline
Initialize &  \makecell[l]{$\mathbb{V}^o_{\text{o\_orderkey}}=\Pi_{\text{o\_orderkey}}(G^o_1(\texttt{orders}))=\{1,3,4,...\}$\\$\mathbb{V}^l_{\text{l\_orderkey}}=\Pi_{\text{l\_orderkey}}(G^l_2(\texttt{lineitem}))=\{2,3,4,...\}$} \\
\hline
Iter 1 &  \makecell[l]{$\mathbb{V}^o_{\text{o\_orderkey}}=\Pi_{\text{o\_orderkey}}(G^o_1\!\downarrow(\texttt{orders}))=\{3,4,...\}$\\$\mathbb{V}^l_{\text{l\_orderkey}}=\Pi_{\text{l\_orderkey}}(G^l_2\!\downarrow(\texttt{lineitem}))=\{3,4,...\}$} \\
\hline
Iter 2 & same as Iter 1 \\
\hline
\end{tabular}
}
\vspace{-3mm}
\end{table}

\subsection{Illustrating algorithm: TPC-H Q4}
We illustrate how Algorithm \ref{alg:opt-no-snapshot} works on Q4. Table~\ref{tab:q4-no-snapshot} details the predicates in each phase.
 In phase 1, \tool pushes the predicate $F_4=$ \texttt{o\_orderpriority}==$v_g$ through $Op_4$,  producing a lineage superset: $F_3 =$ \texttt{True} on the input from $SubOp_3$ and $F_1 =$ \texttt{o\_orderpriority}==$v_g$ on the input from $Op_1$. It continues pushing them down until reaching $G^o_1$ and $G^l_2$ on the input tables.

Next, the algorithm starts pushing up. It constructs a variable $\mathbb{V}^o_{\text{o\_orderkey}}$ to represent values for the \dbfield{o\_orderkey} column (and similarly for other columns) and creates a row-value predicate (\dbfield{o\_orderkey}$\in \mathbb{V}^o_{\text{o\_orderkey}}\wedge\cdots$). 
The same process is performed on the \dbtable{lineitem} table, where $\mathbb{V}^l_{\text{l\_orderkey}}$ represents values from \dbfield{l\_orderkey}. These two sets of values merge in $G_5\!\uparrow$.

In phase 3, at $Op_4$,  \tool pushes down a more selective predicate, which includes the original $F_4$, the pushup predicate $F_4\!\uparrow$,  and the predicate pushed down from $Op_5$. Continuing this pushdown process results in the exchange of set variables $\mathbb{V}^o_{\text{o\_orderkey}}$ and $\mathbb{V}^l_{\text{l\_orderkey}}$, where in the end $\mathbb{V}^o_{\text{o\_orderkey}}$ is used to filter the \dbtable{lineitem} table and $\mathbb{V}^l_{\text{l\_orderkey}}$ is used to filter the \dbtable{orders} table. 

In phase 4, \tool first runs $G^o_1$ on the \dbtable{orders} and $G^l_2$ on the \dbtable{lineitem} to initialize the value sets $\mathbb{V}^o_{\text{o\_orderkey}}$ and $\mathbb{V}^l_{\text{l\_orderkey}}$. In the 1st iteration, it runs $G^o_1\!\downarrow$ and $G^l_2\!\downarrow$, where the sets $\mathbb{V}^o_{\text{o\_orderkey}}$ and $\mathbb{V}^l_{\text{l\_orderkey}}$ make these predicates more selective than the original $G^o_1$ and $G^l_2$, e.g., selecting only lineitems that can be joined with orders rather than all lineitems returned by $G^l_2$. In iteration 2, $G^o_1\!\downarrow$ and $G^l_2\!\downarrow$ select the same rows as in the prior iteration, so \tool stops.
The final lineage contains precisely the orders that satisfy the filter condition and are joinable, resulting in zero false positives (the same for lineitems).

\subsection{Scope and limitations}
\label{sec:no-snapshot-limitation}
This algorithm works best for inner joins and equivalence semi-joins like Q4, usually achieving 0 FPR. However, false positives can still arise when the lineage from a source table cannot be pushed up, as commonly seen in certain anti-joins or non-equal semi-joins. For instance, consider a query (select sum($R.a$) from $R$ where $R.a>$ (select avg($S.b$) from $S$);). For a given output $t_o$, the lineage of $S$ would be the entire table, while the lineage of $R$ consists only of rows where $R.a$ > avg($S.b$). In this case, our algorithm is unable to compute the precise lineage, as the information from table $S$ cannot be pushed up to filter table $R$, resulting in both tables being entirely included. The size of the false positives depends on the selectivity of the operators (e.g., the non-equal semi-join) that fail to yield an equivalent pushdown result to pushing down a row-selection predicate, like $R.a$>avg$(S.b)$ in this case.

\section{Evaluation}\label{sec:eval}
\subsection{Experimental setup}
\subsubsection{Query and Pipeline corpus}
We assessed \tool on both relational queries (TPC-H \cite{tpch}) and real-world data pipelines. TPC-H queries include computationally complex operations such as aggregations, joins, and nested subqueries.  We generated 1GB of data and created indexes on the primary keys for the TPC-H tables.

For data science pipelines, we used 70 real-world data processing pipelines sampled from Jupyter Notebooks on GitHub that use \texttt{Pandas} library~\cite{pandas} to clean, transform, or process data. These pipelines consist of up to 234 operators (in PredTrace syntax), handling input data sizes of up to 299MB and involving up to 5 input tables. Additional UDF details on pipeline specifics are provided in Section \ref{sec:eval_ppl}. Using the notebook-replay technique from AutoSuggest \cite{autosuggest}, we obtained the replay trace, which represents the sequence of executed functions. The longest dataflow path in each notebook was identified as the data processing pipeline, excluding side paths unrelated to intermediate results. These pipelines typically generate cleaned data, which is subsequently saved for further processing or used for visualization. From each pipeline, we randomly selected one output item and used \tool to trace its lineage.

\tool uses the Spark SQL parser and has its own Python parser, so it can translate a SQL query or Pandas pipeline into intermediate operator representations as shown in Table \ref{tab:basic-ops}.
To save intermediate results, \tool modifies the Python code by inserting a ``save'' statement or generates split queries for SQL queries.
It produces the final lineage query in SQL (for TPC-H queries) or Pandas (for pipelines).

We implemented all the algorithms using Python 3.7 with the Z3 \cite{z3} SMT solver. TPC-H-related evaluation was performed on PostgreSQL 12.14, and the data science pipelines were run in-memory with Pandas version 1.1.4. We conducted all the experiments on a server with a 2.4GHz processor and 64GB memory.

\vspace{-1mm}
\subsubsection{Baselines} \label{sec:baselines}
In our experiments on TPC-H, we compared \tool to state-of-the-art lazy lineage systems. We executed all queries on the same TPC-H dataset and PostgreSQL server configuration.

\begin{asparaitem}
\item \textbf{Trace} \cite{cui2003lineage} 
uses a lineage tracing procedure that applies to each input row to determine whether it belongs to the lineage set of the given output data item. It iterates over each input row, applying the transformations to identify all input rows that contribute to the generation of the output data item as lineage.

\item \textbf{GProM} \cite{arab2014generic, niu2017provenance, glavic2009provenance} is a provenance middleware that captures how rows from the input are combined to produce the result, which provides more expressive information than lineage. It rewrites the original query to propagate provenance along with the results, incorporating provenance-specific optimizations in the generation of the rewritten query.  While GProM does not handle correlated subqueries,  we leveraged the technique it employs \cite{glavic2009provenance} using only standard query optimizations provided by PostgreSQL to augment its coverage.  As GProM produces a provenance query that returns all output rows with their provenance from source tables, for a fair comparison, we optimized the query by adding a filter to select only $t_o$ and a projection on the provenance columns.

\item \textbf{Panda} \cite{ikeda2013logical} uses a logical-provenance specification language with {\it attribute mappings} and {\it filters}. It generates \textit{attribute mappings} for SELECT clause attributes and input attributes, as well as \textit{filters} for WHERE clause conditions that apply to each source table. If the minimal lineage cannot be obtained, Panda uses {\it augmentations} to encode more precise lineage information.

\end{asparaitem}

\begin{figure}
    \centering
  \includegraphics[scale=0.37]{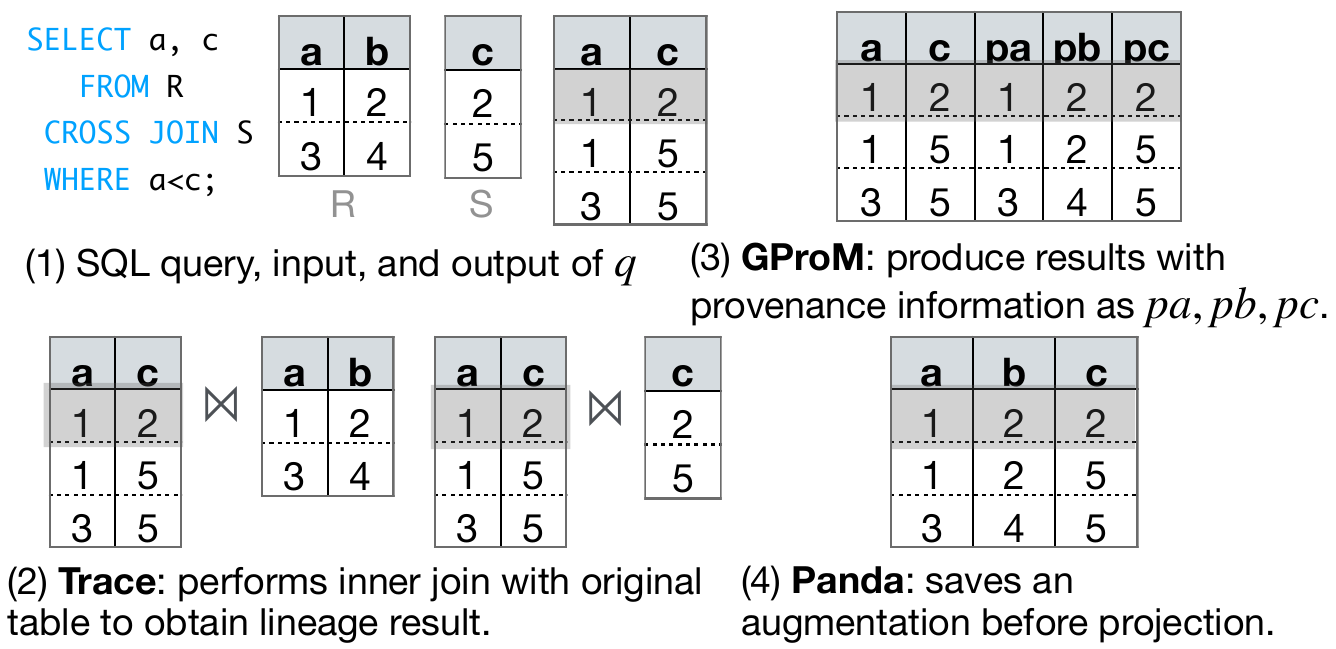}
    \vspace{-4mm}
\caption{\textnormal{Lineage example: baseline approaches.}}
    \vspace{-5mm}
    \label{fig:baseline}
\end{figure}

\begin{figure*}[]
\begin{minipage}{0.7\textwidth}
     \centering
		\includegraphics[width=\linewidth]{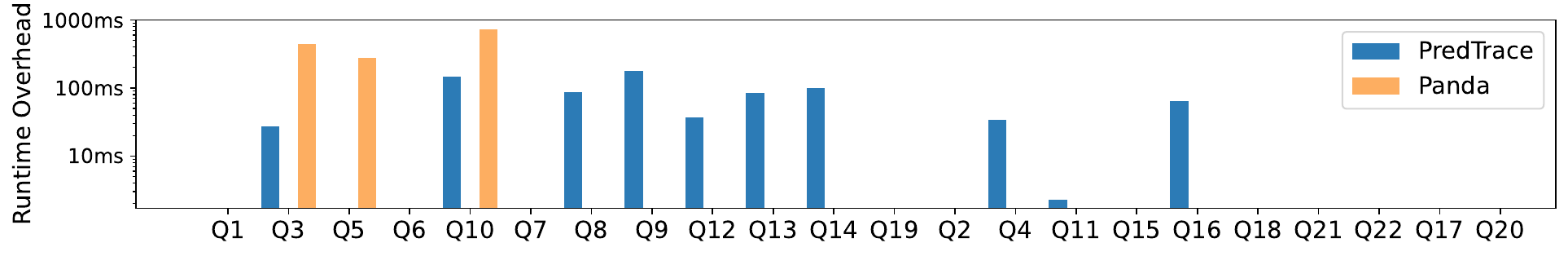}
			\vspace{-9mm}
		\caption{\textnormal{Runtime overhead for TPC-H (log scale).}}
		\label{fig:overhead}
\end{minipage}
\begin{minipage}{0.26\textwidth}
\centering
\small
  	\begin{tabular}[b]{l|ccc}  	
  \toprule
	{\bf Method	}& {\bf min} & {\bf max} &{\bf avg}\\ 
  \midrule
		 \tool	& 0 &  179.7& 34.7\\ 
     		\hline
     	 Panda		& 0 & 721.6  &482.1\\ 
		\bottomrule
	\end{tabular}
 \vspace{-4mm}
    \caption{\textnormal{Overhead statistics (ms).}}
    \label{tbl:statistic}
\end{minipage}
\vspace{-4.7mm}
\end{figure*}

\begin{figure*}[]
\begin{minipage}{0.7\textwidth}
     \centering
		\includegraphics[width=\linewidth]{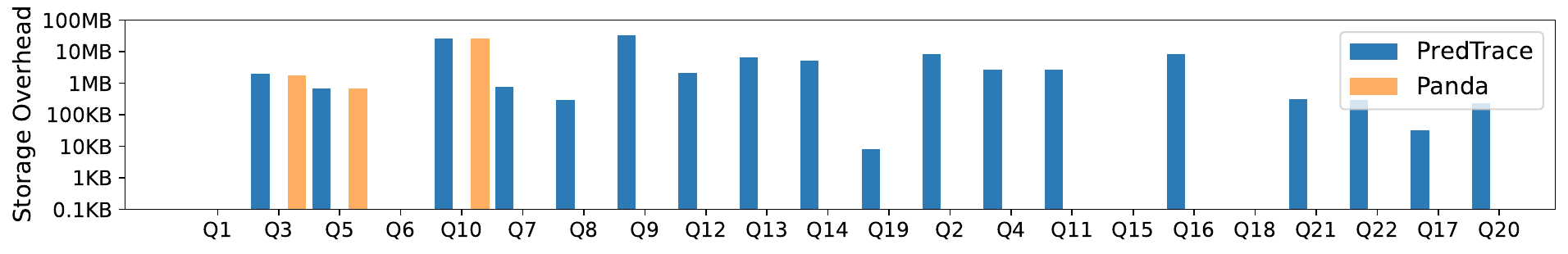}
			\vspace{-9mm}
   \caption{\textnormal{Storage overhead for TPC-H (log scale).}}
		\label{fig:storage}
\end{minipage}
\begin{minipage}{0.26\textwidth}
\centering
\small
  	\begin{tabular}[b]{l|ccc}  	
  \toprule
	{\bf Method	}& {\bf min} & {\bf max} &{\bf avg}\\ 
  \midrule
		 \tool	& 0 & 33792 & 4531\\ 
     		\hline
     	 Panda	& 0 & 25600  & 5613\\ 
		\bottomrule
	\end{tabular}
 \vspace{-4mm}
    \caption{\textnormal{Storage statistics (KB).}}
    \label{tbl:storage_statistic}
\end{minipage}
\vspace{-4.7mm}
\end{figure*}

\begin{figure*}[]
\begin{minipage}{0.7\textwidth}
     \centering
		\includegraphics[width=\linewidth]{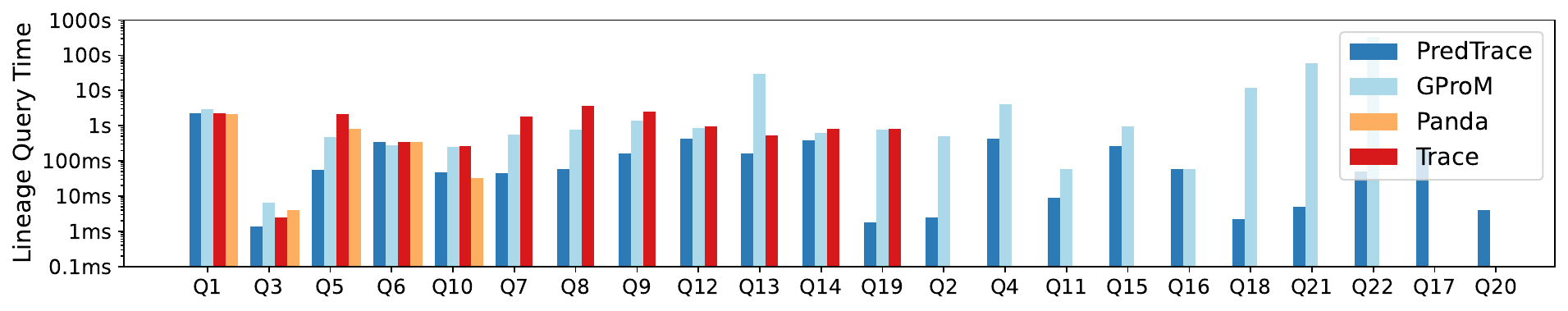}
			\vspace{-9mm}
   \caption{\textnormal{Lineage query time for TPC-H (log scale).}}
		\label{fig:query}
\end{minipage}
\begin{minipage}{0.28\textwidth}
\centering
\small
  	\begin{tabular}[b]{l|ccc}  	
  \toprule
	{\bf Method	}& {\bf min} & {\bf max} &{\bf avg}\\ 
  \midrule
		 \tool	& 1.4 & 2253.9 & 226.6\\ 
   		\hline
            GProM & 6.5 & 330182.0& 22272.2\\
            \hline
       	 Panda	& 4.1 & 2168.0 & 677.7\\ 
		\hline
       	 Trace	& 2.4 &  3706.5 & 1343.5\\ 
		\bottomrule
	\end{tabular}
     \vspace{-6mm}
    \caption{\textnormal{Query time statistics (ms).}}
    \label{tbl:query_statistic}
\end{minipage}
\vspace{-4mm}
\end{figure*}

To better explain the baseline approaches, let's consider an example query $q$ shown in Figure \ref{fig:baseline}. The input relations are $R$ = \{(1, 2), (3, 4)\} and $S$ = \{(2), (5)\}, with schemas $R=(a, b)$ and $S=(c)$, and the SQL query of $q$ is (select $a, c$ from $R$ cross join $S$ where $a<c$;).
To trace the lineage for the output row $t_o$ = (1, 2) (marked in grey in Figure \ref{fig:baseline}), \textbf{Trace} method applies a filter $F = (a = 1 \wedge c = 2)$ to identify $t_o$ in $q$'s output and then uses it to select the corresponding lineage from each input table using inner joins.
\textbf{GProM} reveals how each output is derived by rewriting the original query as $q^+$: (select $a$, $c$, $a$ as $pa$, $b$ as $pb$, $c$ as $pc$ from $R$ cross join $S$ where $a<c$;). To obtain the lineage of $t_o$, a filter $F = (a=1 \wedge c=2)$ is added to $q^+$, and the results are projected onto the provenance columns $pa, pb$ and $pc$, resulting in running (select $a$ as $pa$, $b$ as $pb$, $c$ as $pc$ from $R$ cross join $S$ where $a=1$ and $c=2$;).
\textbf{Panda} saves an augmentation before projecting columns $a$ and $c$, to maintain the lineage information. It derives the attribute mapping $M=$\{$O.a \leftrightarrow R.a, O.c \leftrightarrow S.c$\} and the filter $F=\{\sigma_{(R.a=1)} \wedge \sigma_{(R.b=2)}, \sigma_{(S.c=2)}\}$ to select the lineage set based on the augmentation.

For data science pipelines, none of the existing lazy methods handle the wide range of operators/UDFs that these pipelines involve. Instead, they require users to manually specify logical provenance \cite{cui2003lineage, ikeda2013logical}. We constructed a baseline to eagerly track the lineage by adding an extra column to each dataframe to store the table ID and primary key of the input rows contributing to each output.

\vspace{-2mm}
\subsection{Evaluation on TPC-H}\label{sec:eval-tpch}

Our experiments aim to demonstrate that \tool (1) incurs minimal time and storage overhead during query execution to materialize intermediate results, (2) supports efficient lineage querying, and (3) effectively reduces both time and storage overhead through intermediate result optimization. In all figures, queries are sorted by the order of support in the baseline methods, with query coverage discussed in Table \ref{tbl:coverage}. Among all baselines, \tool is the only method that successfully supports all 22 queries.

\paragraph{Query execution overhead.}
\tool may incur additional runtime and storage overhead during query execution by materializing intermediate results. Figure~\ref{fig:overhead}, \ref{tbl:statistic} show the runtime overhead compared to the original TPC-H queries, and Figure~\ref{fig:storage}, \ref{tbl:storage_statistic} show the storage overhead. On average, \tool adds only 34.7 ms to the query runtime and requires an additional 4.5 MB of storage, which is relatively minimal considering that the average runtime for the original TPC-H queries in our database is 336,067 ms, with a database size of 1 GB.
Among the 22 queries, 4 of them (queries 1, 6, 15, and 18) do not save any intermediate results, thus incurring no overhead. Queries 2, 3, 4, 5, 7, 8, 9, 11, 12, 13, 14, 16, 17, 19, 21, and 22 save one intermediate result, and queries 10 and 20 save two intermediate results, with a maximum runtime increase of 179.7 ms and a 34 MB intermediate result.

Panda has limited support for query types (only queries 1, 3, 5, 6, and 10). When augmentation is required for queries 3, 5, and 10, Panda incurs additional time and storage overhead during query execution, which is comparable to \tool because the augmentation preserves information similar to the intermediate results.

GProM and Trace do not incur any runtime or storage overhead since the query runtime is unmodified. However, they are orders of magnitude slower when querying lineage, as demonstrated later.

\begin{table}[tp]
  \centering
  \footnotesize
  \caption{\textnormal{TPC-H query coverage (database size: 1GB).}}
  \vspace{-2mm}
  \label{tbl:coverage}
  \begin{tabular}{cc}
    \toprule
    \textbf{Approach} & \textbf{\# queries supported} \\
    \midrule
    {\bf \tool} & {\bf 22 (/22) } \\
   GProM & 20 (/22), except Q17, 20 over 6hr  \\
    Panda & 5 (/22), can only handle single SELECT blocks  \\
    Trace & 12 (/22), can only handle non-nested queries \\
    \hline
    {\bf \tool-iterative} & {\bf 22 (/22)} \\
    \bottomrule
  \end{tabular}
  \vspace{-2mm}
\end{table}

\paragraph{Lineage query time.}
In Figure \ref{fig:query}, we compared the lineage querying time. \tool's lineage querying time includes the time to execute predicate $F_i$ on the materialized intermediate result to replace variables in $F_i^{row}$ and the time to run the pushed-down predicates on the source tables. On average, lineage query takes 226.6 ms (up to 2253.9 ms), which is significantly better than other baselines. The lineage query time for \tool is generally much faster than the runtime of the original queries. This is because, with predicate pushdown, the lineage query only involves executing predicates on intermediate results and source tables, without needing to perform operations like joins in the original query.

On average, GProM takes 22s for lineage querying (up to 330s, except for Q17 and Q20 which take over 6hrs). It demonstrates comparable performance to \tool in 5 out of the 22 queries but incurs higher time overhead for more complex queries. In contrast, \tool exhibits a lower lineage query time, primarily influenced by the size of intermediate results and source tables as it is performing scans on them.

Compared to the other two baselines, \tool consistently performs significantly better in all queries. This is because Trace typically issues a complicated query, sometimes re-executing the original query to compute the lineage. Panda does not ensure the storage of the primary key in the augmentation, resulting in slower retrieval of lineage rows from the source tables, as it cannot leverage the primary key index on the source tables.

\paragraph{Effectiveness of intermediate result optimization.}
We assessed the effectiveness of the intermediate result optimization in Section \ref{sec:opt}. While the local optimization of column projection is a straightforward approach applicable to any query, we focused on the global optimization of selecting intermediate results with minimal sizes.

Global optimization is beneficial when multiple joins are present and deferring materialization yields the same lineage. Consequently, only queries 3, 5, 7, and 19 benefit from this optimization among the 18 queries that materialize intermediate results. Table \ref{tbl:optimize} compares the number of rows and the sizes of the naive and optimized intermediate results, and the lineage querying time.

For the 4 queries affected, this optimization brings a significant reduction in intermediate result sizes, ranging from 95\% to 99\%, and achieves a 2$\times$ to 270$\times$ speedup in lineage querying.

\begin{table}[tp]
\caption{\textnormal{Comparison of storage overhead and lineage query time.}} \label{tbl:optimize}
\vspace{-4mm}
\centering
\footnotesize
\begin{tabular}{l|c|c|c|c|c|c}
\toprule
& \multicolumn{3}{l}{\tool-N} & \multicolumn{3}{l}{\tool-O} \\
\midrule
 Q\_id     & \# of rows        & size        & \makecell[c]{lineage \\ query time} & \# of rows        & size   &   \makecell[c]{lineage \\query time}  \\
      \hline
3     &         147,126           &         44 MB    &0.73s &          30, 519   &  1984 KB     &0.43s \\
5     &          182,183         &      138 MB       &1.67s&         7,243           &      664 KB    &0.27s   \\
7     &        6,001,215           &      5.2 GB       &100.40s&            5,924        &       744 KB   & 0.37s  \\
19    &            6,001,215       &        1676 MB      &30.62s&            121        &      8.2 KB     &0.42s  \\
\bottomrule
\end{tabular}
\vspace{-2mm}
\end{table}

\paragraph{Logical inference time.} 
All systems are efficient in logical inference, which includes the time to push down predicates to determine the lineage query. This time depends solely on the pipeline complexity and remains unaffected by the size of the database. For all approaches, the logical lineage inference time consistently stays below one second.

\begin{table}[tp]
\caption{\textnormal{FPR with naive pushdown and iterative refinement.}} \label{tbl:fpr}
\vspace{-4mm}
\centering
\footnotesize
\begin{tabular}{c|c|c|c|c|c|c}
\toprule
& Q2,3,4,5,7,8,9,10,20 & Q11 & Q12 & Q13 & Q14 & Q16 \\
\hline
Naive & 99\% & 96\% & 98\% & 1\% & 49\% & 99\% \\
Iterative & 0\% & 0\% & 0\% & 0\% & 0\% & 10\% \\
\hline
& Q1,6,15,18 & Q17 & Q19 & Q21 & Q22 & \textbf{avg} \\
\hline
Naive & 0\% & 99\% & 97\% & 99\% & 67\% & 70.7\% \\
Iterative & 0\% & 12\% & 0\% & 80\% & 43\% & 6.6\%\\ 
\bottomrule
\end{tabular}
\vspace{-2mm}
\end{table}

\begin{figure}
    \centering
    \includegraphics[scale= 0.25]{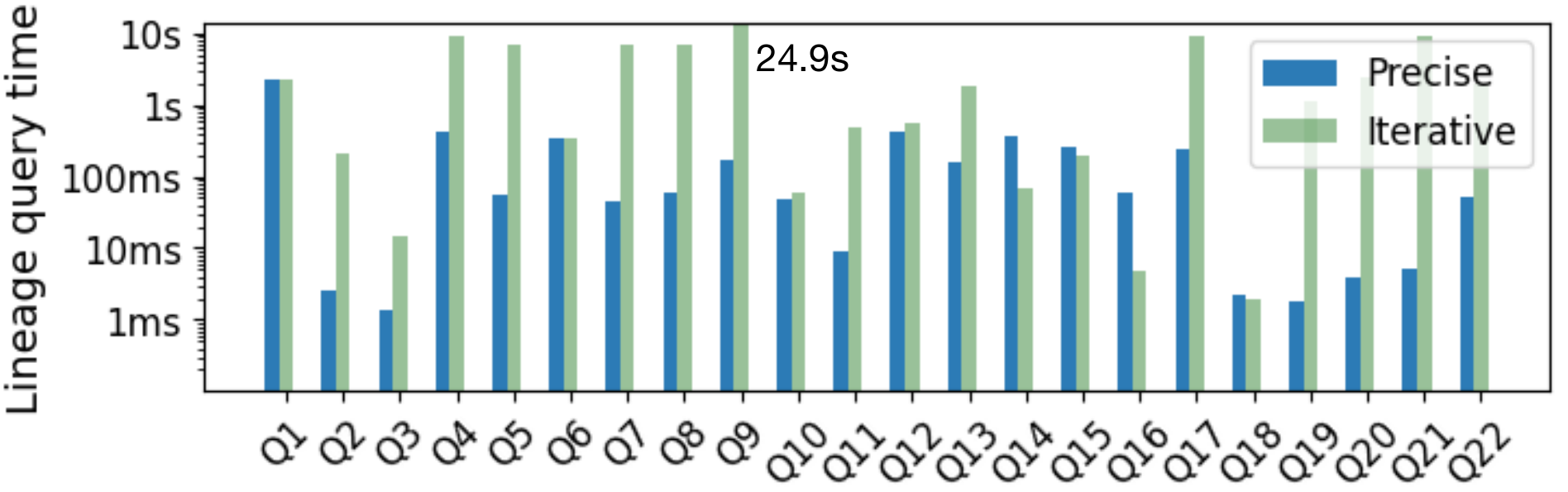}
    \vspace{-4mm}
    \caption{\textnormal{Lineage query time: with vs. without saving intermediate results.}}
    \label{fig:iter_query}
    \vspace{-2mm}
\end{figure}

\subsection{Without saving intermediate results}
\label{sec:eval-no-inter}
We evaluated our lineage solution without using intermediate results, which we denote as \textit{Iterative}. We compared the false positive rate (FPR) with a naive pushdown approach, where we simply push down a row-selection predicate and apply it at the source tables as is (without the subsequent pushup and iterative filtering phases). The results are shown in Table~\ref{tbl:fpr}.
With iterative refinement, \textit{Iterative} achieves a significantly lower FPR (6.6\% on average compared to 70.7\%), including a 0 FPR for 18 queries. Exceptions are queries 16, 17, 21, and 22, which contain anti-joins and non-equal semi-joins where the lineage information fails to be pushed up, as discussed in Section~\ref{sec:no-snapshot-limitation}. The false positives include: 1) rows from the outer table that are not properly filtered out by the anti-join/non-equal semi-join, and 2) rows from the inner table that match the outer table rows flowing in before the anti-join (where the precise lineage should be an empty set).

We presented the lineage querying time in Figure~\ref{fig:iter_query}. \textit{Iterative} has a higher lineage querying time (3852.1 ms on average) compared to saving intermediate results, referred to as \textit{Precise} (226.6 ms on average). This is because the iterative refinement scans all tables at each iteration. However, it remains comparable to or better than other baselines listed in Figure \ref{tbl:query_statistic}.
For queries containing inner-joins and semi-joins, the number of iterations often increases linearly with the number of joins, making it slower for queries with large join graphs.

\subsection{Comparison to lineage tracking systems.} 
We compared \tool to the state-of-the-art eager lineage tracking system, SMOKE~\cite{psallidas2018smoke}, which tightly integrates lineage tracking by creating lineage indexes during query execution. In contrast, \tool adopts a lazy inference approach, inferring lineage through predicate pushdown and only storing intermediate results when precise lineage cannot be obtained. This approach offers \tool greater adaptability, as it avoids the need to change operator implementations or rely on specific data systems.

For the performance comparison, we utilized the reported results from a subset of TPC-H queries in \cite{psallidas2018smoke}, since SMOKE is not open source. These results reflect SMOKE's optimal lineage capture performance with all optimizations enabled. Across the four evaluated queries (Q1, 3, 10, and 12), SMOKE incurs an average query execution overhead of 14.5\%, whereas \tool achieves a smaller average overhead of 10.35\%\footnote{A direct time comparison in milliseconds is not feasible, as the queries were executed in different DBMS environments with varying runtimes.}.
However, SMOKE demonstrates superior efficiency in lineage querying (under 100 ms), primarily due to the eager creation of indexes during tracking, which enables fast index scans. In contrast, \tool conducts a full source table scan during lineage queries. We acknowledge this limitation and plan to explore indexing support in future work.

\begin{table}[tp]
\footnotesize
  \caption{\textnormal{Summary of embedded UDFs in data science pipelines.}}
  \vspace{-4mm}
  \label{tab:udf}
  \begin{centering}
  \begin{tabular}{lll}
  \toprule
  {\bf Category}  & {\bf UDF usage}& \makecell[l]{\textbf{\# of ppls} \\\textbf{containing}}\\
  \midrule
  Selection-predicate  & \makecell[l]{dataframe filter, dropna} & 42 \\
  Join-predicate  & inner/outer join & 22 \\
  Row-transform  & \makecell[l]{assignment, column rename, change type, \\lambda func, split, get dummies, etc.
  }&  59 \\
  Aggregation  & groupby, aggregation, pivot & 38\\
  Compare  & sort, top-k  &  22\\
  SubQuery & \makecell[l]{iterate rows in a table to output a new row\\ (e.g. data imputation)} & 8\\
  GroupedMap  &  \makecell[l]{transform grouped sub-tables\\ (e.g. customized normalization)}&  9\\
  \bottomrule
\end{tabular}
\end{centering}
 \vspace{-5mm}
\end{table}

\subsection{Evaluation on data science pipelines}\label{sec:eval_ppl}

In our evaluation, the real-world data pipelines cover a wide range of non-relational operators such as pivot, unpivot, and operators with non-trivial UDFs. 
\tool provides robust support for UDFs embedded in operators, accommodating various \texttt{Pandas} APIs \cite{pandas} that facilitate the use of functions as input parameters through \texttt{apply, transform, query, groupby}, etc. The functions are defined using lambda in Python syntax, encompassing statements like assignment, \texttt{for}, \texttt{if}, and \texttt{return}. We present the usage of UDFs across these 70 pipelines in Table \ref{tab:udf}.

\begin{figure}[tp]
    \centering
    \includegraphics[scale= 0.22]{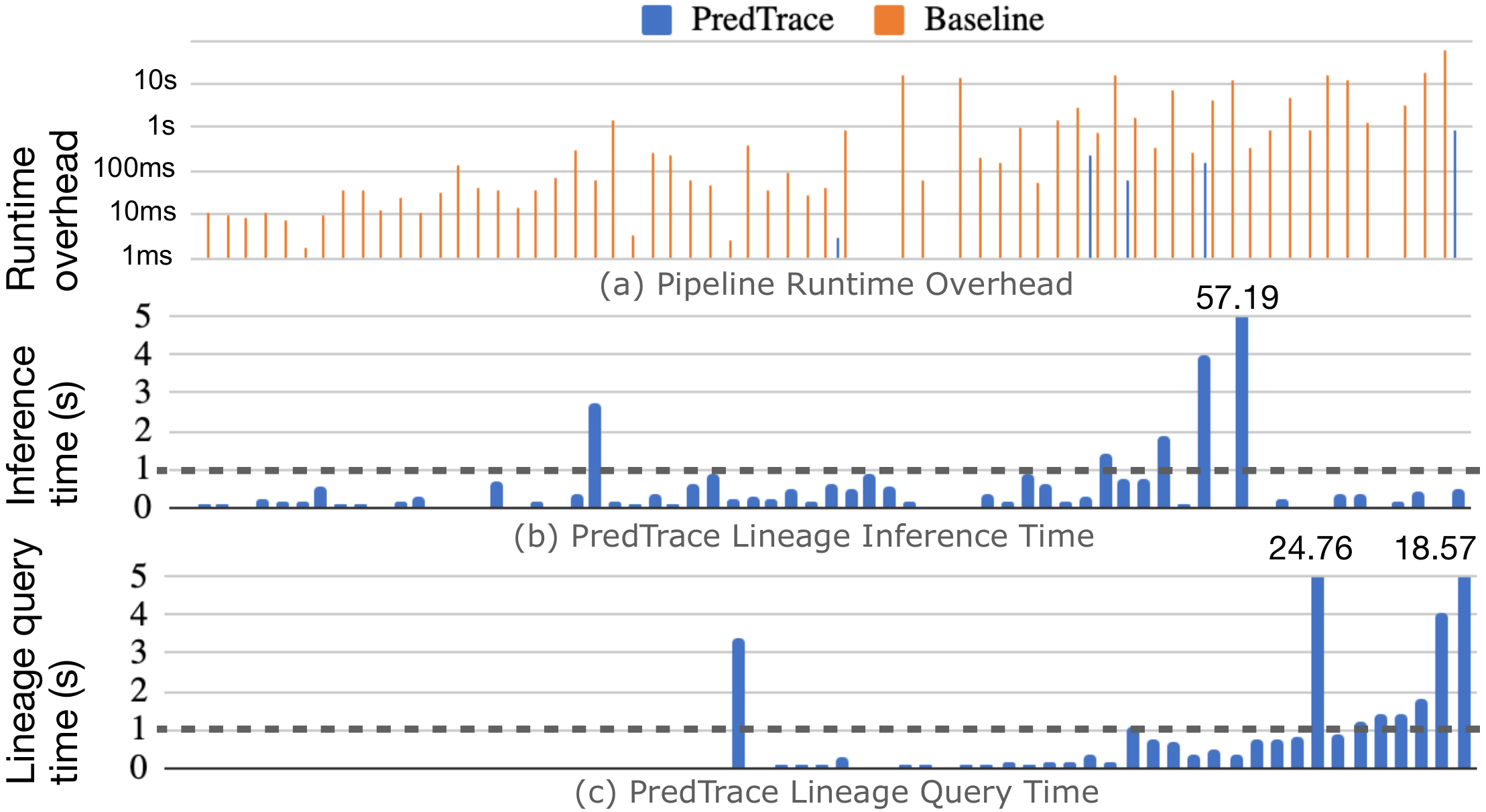}
    \vspace{-2mm}
    \caption{\textnormal{Pipeline runtime increased (log scale), lineage inference time, and lineage query time, sorted by \# input rows.}}
    \label{fig:ppl_details}
    \vspace{-3mm}
\end{figure}

We compared \tool with a lineage tracking baseline discussed in Section \ref{sec:baselines}. We first evaluated the runtime overhead of each pipeline, as shown in Figure \ref{fig:ppl_details} (a).
\tool significantly outperformed the baseline and incurred no execution overhead in 65 pipelines. For the remaining 5 pipelines, it saved only one intermediate result, with runtime overhead consistently under 1 second. Larger intermediate results would have led to greater runtime overhead.

The baseline approach demonstrates lower time overhead in cases where the pipelines consist of simple operators and small datasets. However, as the number of operators and dataset sizes increase, the lineage tracking method incurs a higher time overhead to maintain and compute the lineage information. Our evaluation shows that the baseline overhead can reach up to 60 seconds, which is over 10$\times$ the pipeline execution time in these cases.

We also measured the logical inference time and lineage query time of \tool. Figure \ref{fig:ppl_details}(b) shows that the lineage inference time is mainly influenced by the predicate pushdown overhead of MagicPush. In most pipelines, the lineage inference time remains below 1 second. The slowest pipeline involves a UDF with 14 branch conditions, causing MagicPush to take longer to search and verify the correct pushed-down predicate $G$. The second slowest pipeline is affected by a Pivot operator with a wide input table, leading to a longer verification time.  For other pipelines exceeding 1 second, the delay is primarily due to the presence of numerous operators.

Regarding lineage query time (Figure \ref{fig:ppl_details} (c)), we observed a strong correlation between the input table size and intermediate result size. When ranking pipelines by the number of input rows, we found that for most pipelines with a small input size (and consequently small intermediate result), the lineage query time consistently remains below 0.1s. Outliers with longer query times are primarily due to UDFs using libraries to process data in the pushed-down predicate. For instance, a type transformation from string to datetime could result in pushed-down predicates like \texttt{pd.to\_datetime(x["Date"]) == '2015-04-15 00:00:00'}, and the slow library function would lead to a higher query time.

\vspace{1mm}
{\it Without saving intermediate results.} We also assessed our iterative refinement method across the pipelines. It does not affect the 65 pipelines where no intermediate results need to be saved. For the remaining 5 pipelines, iterative refinement is able to return the same lineage as using intermediate results, with 0 false positives. It has a comparable lineage querying time to saving intermediate results (an average of 5.7 seconds compared to 5 seconds) because it stops after only two iterations for all pipelines (thus scanning each table three times) and avoids scanning large intermediate results.

\section{Related Work} \label{sec:related}

Buneman et al. \cite{buneman2001and} describe two semantics of data provenance: the influence-contribution semantics (a.k.a. why-provenance) \cite{cui2000tracing}, which identifies the minimum set of data elements responsible for a result, and the copy-contribution semantics (a.k.a. where-provenance) \cite{buneman2001and}, which considers the provenance of individual attribute values. Green et al. \cite{green2007provenance} formalize the notion of how-provenance \cite{hernandez2021computing}, which not only identifies which input tuples contribute but also explains how they contribute, using an appropriate "provenance semiring" in the context of relational algebra.
Another approach called causality \cite{meliou2010complexity} not only captures the contributing elements but also measures the degree of their contribution to the result. \tool focuses on why-provenance.

Existing lineage tracing mainly focuses on relational transformations.
In \cite{cui2000tracing,cui2000practical}, the authors solve the view data lineage problem using inverse queries to trace lineage; however, the drawback is that the lineage results are not expressed in relational algebra, limiting DBMS query optimizations.
To address this, Perm \cite{glavic2009perm} employs query rewrite mechanisms to compute the provenance of an SQL query and extends it to support nested queries \cite{glavic2009provenance}. GProM \cite{arab2014generic} serves as a versatile DBMS middleware capable of handling various provenance types across queries, updates, and transactions. Furthermore, \cite{niu2017provenance} introduces provenance-aware optimizations aimed at enhancing the performance of instrumented queries in the GProM system.
Trio \cite{widom2004trio} extends the relational model with uncertainty and makes lineage an integral part of the data model. SMOKE \cite{psallidas2018smoke} is a database engine that eagerly materializes data structures during query execution to achieve interactive speed for both lineage capture and lineage query execution.

Moreover, data-intensive scalable computing systems like Apache Spark \cite{zaharia2010spark} and the Titan extension \cite{interlandi2015titian} have been developed to provide efficient fine-grained forward and backward lineage capture and query support. BigDebug \cite{gulzar2016bigdebug}, built on Titan, enables users to set breakpoints and on-demand watchpoints, and locate the record causing a crash. However, many of these systems are tightly integrated with underlying platforms and do not adequately cover UDFs in data science semantics.


Lineage tracing has also been explored in data-oriented workflows. Amsterdamer et al. \cite{amsterdamer2011putting} propose a lineage model using Pig Latin \cite{olston2008pig} to express workflow module functionality, but it's closely tied to Pig Latin operations and may not suit general operators.
For more general operators, \cite{cui2003lineage} considers general transformations but doesn't delve into lineage properties such as correctness and minimality.
Another approach in \cite{ikeda2013logical} supports arbitrary transformations using \textit{attribute mapping} and \textit{filtering} but requires the manual specification of filters and attribute mappings for non-relational operators.
Vizier \cite{brachmann2020your} is a notebook-style data exploration system that aids analysts in building and refining data pipelines using provenance tracking. In contrast, \tool provides a lightweight and comprehensive solution based on predicate pushdown. PROVision \cite{zheng2019fine} reconstructs fine-grained provenance for ETL workflows and optimizes by pushing down a semijoin to focus on specific output subsets. However, it does not directly obtain lineage from source tables via pushdown.

The idea of leveraging a predicate on one table to filter other tables, as used by our iterative method, has been explored for query optimization~\cite{data-induced-predicate, bitvector}. However, these works only handle relational SPJA queries and push predicates only to tables where composite indexes are available and inexpensive. We, instead, push predicates to all tables and iteratively refine them to reduce false positives.

\section{Conclusion}
In this paper, we present \tool, a tool that utilizes predicate pushdown to infer lineage. \tool supports both saving intermediate results for precise lineage and operating without intermediate results for lineage supersets. Experiments show that \tool achieves low pipeline runtime overhead, efficient lineage querying, and high query coverage.

\balance 
\clearpage
\bibliographystyle{ACM-Reference-Format}
\bibliography{ref}


\begin{thebibliography}{39}


\ifx \showCODEN    \undefined \def \showCODEN     #1{\unskip}     \fi
\ifx \showDOI      \undefined \def \showDOI       #1{#1}\fi
\ifx \showISBNx    \undefined \def \showISBNx     #1{\unskip}     \fi
\ifx \showISBNxiii \undefined \def \showISBNxiii  #1{\unskip}     \fi
\ifx \showISSN     \undefined \def \showISSN      #1{\unskip}     \fi
\ifx \showLCCN     \undefined \def \showLCCN      #1{\unskip}     \fi
\ifx \shownote     \undefined \def \shownote      #1{#1}          \fi
\ifx \showarticletitle \undefined \def \showarticletitle #1{#1}   \fi
\ifx \showURL      \undefined \def \showURL       {\relax}        \fi
\providecommand\bibfield[2]{#2}
\providecommand\bibinfo[2]{#2}
\providecommand\natexlab[1]{#1}
\providecommand\showeprint[2][]{arXiv:#2}

\bibitem[\protect\citeauthoryear{??}{mon}{[n.d.]}]%
        {monotonic-query}
 \bibinfo{year}{[n.d.]}\natexlab{}.
\newblock \bibinfo{title}{Monotonic query.}
\newblock \bibinfo{howpublished}{\url{https://en.wikipedia.org/wiki/Monotonic_query}}.
\newblock


\bibitem[\protect\citeauthoryear{??}{pan}{[n.d.]}]%
        {pandas}
 \bibinfo{year}{[n.d.]}\natexlab{}.
\newblock \bibinfo{title}{Pandas: data analysis and manipulation library in Python}.
\newblock \bibinfo{howpublished}{\url{https://pandas.pydata.org/}}.
\newblock


\bibitem[\protect\citeauthoryear{??}{z3}{[n.d.]}]%
        {z3}
 \bibinfo{year}{[n.d.]}\natexlab{}.
\newblock \bibinfo{booktitle}{\emph{The Z3 theorem prover.}}
\newblock
\newblock
\shownote{\url{https://github.com/Z3Prover/z3}.}


\bibitem[\protect\citeauthoryear{Amsterdamer, Davidson, Deutch, Milo, Stoyanovich, and Tannen}{Amsterdamer et~al\mbox{.}}{2011}]%
        {amsterdamer2011putting}
\bibfield{author}{\bibinfo{person}{Yael Amsterdamer}, \bibinfo{person}{Susan~B Davidson}, \bibinfo{person}{Daniel Deutch}, \bibinfo{person}{Tova Milo}, \bibinfo{person}{Julia Stoyanovich}, {and} \bibinfo{person}{Val Tannen}.} \bibinfo{year}{2011}\natexlab{}.
\newblock \showarticletitle{Putting lipstick on pig: Enabling database-style workflow provenance}.
\newblock \bibinfo{journal}{\emph{arXiv preprint arXiv:1201.0231}} (\bibinfo{year}{2011}).
\newblock


\bibitem[\protect\citeauthoryear{Arab, Gawlick, Radhakrishnan, Guo, and Glavic}{Arab et~al\mbox{.}}{2014}]%
        {arab2014generic}
\bibfield{author}{\bibinfo{person}{Bahareh Arab}, \bibinfo{person}{Dieter Gawlick}, \bibinfo{person}{Venkatesh Radhakrishnan}, \bibinfo{person}{Hao Guo}, {and} \bibinfo{person}{Boris Glavic}.} \bibinfo{year}{2014}\natexlab{}.
\newblock \bibinfo{title}{A generic provenance middleware for database queries, updates, and transactions}.
\newblock
\newblock


\bibitem[\protect\citeauthoryear{Bhagwat, Chiticariu, Tan, and Vijayvargiya}{Bhagwat et~al\mbox{.}}{2005}]%
        {bhagwat2005annotation}
\bibfield{author}{\bibinfo{person}{Deepavali Bhagwat}, \bibinfo{person}{Laura Chiticariu}, \bibinfo{person}{Wang-Chiew Tan}, {and} \bibinfo{person}{Gaurav Vijayvargiya}.} \bibinfo{year}{2005}\natexlab{}.
\newblock \showarticletitle{An annotation management system for relational databases}.
\newblock \bibinfo{journal}{\emph{The VLDB Journal}}  \bibinfo{volume}{14} (\bibinfo{year}{2005}), \bibinfo{pages}{373--396}.
\newblock


\bibitem[\protect\citeauthoryear{Brachmann and Spoth}{Brachmann and Spoth}{2020}]%
        {brachmann2020your}
\bibfield{author}{\bibinfo{person}{Michael Brachmann} {and} \bibinfo{person}{William Spoth}.} \bibinfo{year}{2020}\natexlab{}.
\newblock \showarticletitle{Your notebook is not crumby enough, REPLace it}. In \bibinfo{booktitle}{\emph{Conference on Innovative Data Systems Research (CIDR)}}.
\newblock


\bibitem[\protect\citeauthoryear{Buneman, Khanna, and Wang-Chiew}{Buneman et~al\mbox{.}}{2001}]%
        {buneman2001and}
\bibfield{author}{\bibinfo{person}{Peter Buneman}, \bibinfo{person}{Sanjeev Khanna}, {and} \bibinfo{person}{Tan Wang-Chiew}.} \bibinfo{year}{2001}\natexlab{}.
\newblock \showarticletitle{Why and where: A characterization of data provenance}. In \bibinfo{booktitle}{\emph{Database Theory—ICDT 2001: 8th International Conference London, UK, January 4--6, 2001 Proceedings 8}}. Springer, \bibinfo{pages}{316--330}.
\newblock


\bibitem[\protect\citeauthoryear{Council}{Council}{1992}]%
        {tpch}
\bibfield{author}{\bibinfo{person}{Transaction Processing~Performance Council}.} \bibinfo{year}{1992}\natexlab{}.
\newblock \bibinfo{title}{TPC Benchmark H (TPC-H)}.
\newblock \bibinfo{howpublished}{\url{http://www.tpc.org/tpch/}}.
\newblock


\bibitem[\protect\citeauthoryear{Cui and Widom}{Cui and Widom}{2000}]%
        {cui2000practical}
\bibfield{author}{\bibinfo{person}{Yingwei Cui} {and} \bibinfo{person}{Jennifer Widom}.} \bibinfo{year}{2000}\natexlab{}.
\newblock \showarticletitle{Practical lineage tracing in data warehouses}. In \bibinfo{booktitle}{\emph{Proceedings of 16th International Conference on Data Engineering (Cat. No. 00CB37073)}}. IEEE, \bibinfo{pages}{367--378}.
\newblock


\bibitem[\protect\citeauthoryear{Cui and Widom}{Cui and Widom}{2003}]%
        {cui2003lineage}
\bibfield{author}{\bibinfo{person}{Yingwei Cui} {and} \bibinfo{person}{Jennifer Widom}.} \bibinfo{year}{2003}\natexlab{}.
\newblock \showarticletitle{Lineage tracing for general data warehouse transformations}.
\newblock \bibinfo{journal}{\emph{the VLDB Journal}} \bibinfo{volume}{12}, \bibinfo{number}{1} (\bibinfo{year}{2003}), \bibinfo{pages}{41--58}.
\newblock


\bibitem[\protect\citeauthoryear{Cui, Widom, and Wiener}{Cui et~al\mbox{.}}{2000}]%
        {cui2000tracing}
\bibfield{author}{\bibinfo{person}{Yingwei Cui}, \bibinfo{person}{Jennifer Widom}, {and} \bibinfo{person}{Janet~L Wiener}.} \bibinfo{year}{2000}\natexlab{}.
\newblock \showarticletitle{Tracing the lineage of view data in a warehousing environment}.
\newblock \bibinfo{journal}{\emph{ACM Transactions on Database Systems (TODS)}} \bibinfo{volume}{25}, \bibinfo{number}{2} (\bibinfo{year}{2000}), \bibinfo{pages}{179--227}.
\newblock


\bibitem[\protect\citeauthoryear{Ding, Chaudhuri, and Narasayya}{Ding et~al\mbox{.}}{2020}]%
        {bitvector}
\bibfield{author}{\bibinfo{person}{Bailu Ding}, \bibinfo{person}{Surajit Chaudhuri}, {and} \bibinfo{person}{Vivek Narasayya}.} \bibinfo{year}{2020}\natexlab{}.
\newblock \showarticletitle{Bitvector-Aware Query Optimization for Decision Support Queries}. In \bibinfo{booktitle}{\emph{Proceedings of the 2020 ACM SIGMOD International Conference on Management of Data}}. \bibinfo{pages}{2011–2026}.
\newblock


\bibitem[\protect\citeauthoryear{Earley}{Earley}{2015}]%
        {earley2015data}
\bibfield{author}{\bibinfo{person}{Christine~E Earley}.} \bibinfo{year}{2015}\natexlab{}.
\newblock \showarticletitle{Data analytics in auditing: Opportunities and challenges}.
\newblock \bibinfo{journal}{\emph{Business Horizons}} \bibinfo{volume}{58}, \bibinfo{number}{5} (\bibinfo{year}{2015}), \bibinfo{pages}{493--500}.
\newblock


\bibitem[\protect\citeauthoryear{Gershon, Eick, and Card}{Gershon et~al\mbox{.}}{1998}]%
        {gershon1998information}
\bibfield{author}{\bibinfo{person}{Nahum Gershon}, \bibinfo{person}{Stephen~G Eick}, {and} \bibinfo{person}{Stuart Card}.} \bibinfo{year}{1998}\natexlab{}.
\newblock \showarticletitle{Information visualization}.
\newblock \bibinfo{journal}{\emph{interactions}} \bibinfo{volume}{5}, \bibinfo{number}{2} (\bibinfo{year}{1998}), \bibinfo{pages}{9--15}.
\newblock


\bibitem[\protect\citeauthoryear{Glavic and Alonso}{Glavic and Alonso}{2009a}]%
        {glavic2009perm}
\bibfield{author}{\bibinfo{person}{Boris Glavic} {and} \bibinfo{person}{Gustavo Alonso}.} \bibinfo{year}{2009}\natexlab{a}.
\newblock \showarticletitle{Perm: Processing provenance and data on the same data model through query rewriting}. In \bibinfo{booktitle}{\emph{2009 IEEE 25th International Conference on Data Engineering}}. IEEE, \bibinfo{pages}{174--185}.
\newblock


\bibitem[\protect\citeauthoryear{Glavic and Alonso}{Glavic and Alonso}{2009b}]%
        {glavic2009provenance}
\bibfield{author}{\bibinfo{person}{Boris Glavic} {and} \bibinfo{person}{Gustavo Alonso}.} \bibinfo{year}{2009}\natexlab{b}.
\newblock \showarticletitle{Provenance for nested subqueries}. In \bibinfo{booktitle}{\emph{Proceedings of the 12th International Conference on Extending Database Technology: Advances in Database Technology}}. \bibinfo{pages}{982--993}.
\newblock


\bibitem[\protect\citeauthoryear{Green, Karvounarakis, and Tannen}{Green et~al\mbox{.}}{2007}]%
        {green2007provenance}
\bibfield{author}{\bibinfo{person}{Todd~J Green}, \bibinfo{person}{Grigoris Karvounarakis}, {and} \bibinfo{person}{Val Tannen}.} \bibinfo{year}{2007}\natexlab{}.
\newblock \showarticletitle{Provenance semirings}. In \bibinfo{booktitle}{\emph{Proceedings of the twenty-sixth ACM SIGMOD-SIGACT-SIGART symposium on Principles of database systems}}. \bibinfo{pages}{31--40}.
\newblock


\bibitem[\protect\citeauthoryear{Gulzar, Interlandi, Yoo, Tetali, Condie, Millstein, and Kim}{Gulzar et~al\mbox{.}}{2016}]%
        {gulzar2016bigdebug}
\bibfield{author}{\bibinfo{person}{Muhammad~Ali Gulzar}, \bibinfo{person}{Matteo Interlandi}, \bibinfo{person}{Seunghyun Yoo}, \bibinfo{person}{Sai~Deep Tetali}, \bibinfo{person}{Tyson Condie}, \bibinfo{person}{Todd Millstein}, {and} \bibinfo{person}{Miryung Kim}.} \bibinfo{year}{2016}\natexlab{}.
\newblock \showarticletitle{Bigdebug: Debugging primitives for interactive big data processing in spark}. In \bibinfo{booktitle}{\emph{Proceedings of the 38th International Conference on Software Engineering}}. \bibinfo{pages}{784--795}.
\newblock


\bibitem[\protect\citeauthoryear{Hellerstein and Stonebraker}{Hellerstein and Stonebraker}{1993}]%
        {hellerstein1993predicate}
\bibfield{author}{\bibinfo{person}{Joseph~M Hellerstein} {and} \bibinfo{person}{Michael Stonebraker}.} \bibinfo{year}{1993}\natexlab{}.
\newblock \showarticletitle{Predicate migration: Optimizing queries with expensive predicates}. In \bibinfo{booktitle}{\emph{Proceedings of the 1993 ACM SIGMOD international conference on Management of data}}. \bibinfo{pages}{267--276}.
\newblock


\bibitem[\protect\citeauthoryear{Hern{\'a}ndez, Gal{\'a}rraga, and Hose}{Hern{\'a}ndez et~al\mbox{.}}{2021}]%
        {hernandez2021computing}
\bibfield{author}{\bibinfo{person}{Daniel Hern{\'a}ndez}, \bibinfo{person}{Luis Gal{\'a}rraga}, {and} \bibinfo{person}{Katja Hose}.} \bibinfo{year}{2021}\natexlab{}.
\newblock \showarticletitle{Computing how-provenance for SparQL queries via query rewriting}.
\newblock \bibinfo{journal}{\emph{Proceedings of the VLDB Endowment}} \bibinfo{volume}{14}, \bibinfo{number}{13} (\bibinfo{year}{2021}), \bibinfo{pages}{3389--3401}.
\newblock


\bibitem[\protect\citeauthoryear{Ikeda, Sarma, and Widom}{Ikeda et~al\mbox{.}}{2013}]%
        {ikeda2013logical}
\bibfield{author}{\bibinfo{person}{Robert Ikeda}, \bibinfo{person}{Akash~Das Sarma}, {and} \bibinfo{person}{Jennifer Widom}.} \bibinfo{year}{2013}\natexlab{}.
\newblock \showarticletitle{Logical provenance in data-oriented workflows?}. In \bibinfo{booktitle}{\emph{2013 IEEE 29th International Conference on Data Engineering (ICDE)}}. IEEE, \bibinfo{pages}{877--888}.
\newblock


\bibitem[\protect\citeauthoryear{Interlandi, Shah, Tetali, Gulzar, Yoo, Kim, Millstein, and Condie}{Interlandi et~al\mbox{.}}{2015}]%
        {interlandi2015titian}
\bibfield{author}{\bibinfo{person}{Matteo Interlandi}, \bibinfo{person}{Kshitij Shah}, \bibinfo{person}{Sai~Deep Tetali}, \bibinfo{person}{Muhammad~Ali Gulzar}, \bibinfo{person}{Seunghyun Yoo}, \bibinfo{person}{Miryung Kim}, \bibinfo{person}{Todd Millstein}, {and} \bibinfo{person}{Tyson Condie}.} \bibinfo{year}{2015}\natexlab{}.
\newblock \showarticletitle{Titian: Data provenance support in spark}. In \bibinfo{booktitle}{\emph{Proceedings of the VLDB Endowment International Conference on Very Large Data Bases}}, Vol.~\bibinfo{volume}{9}. NIH Public Access, \bibinfo{pages}{216}.
\newblock


\bibitem[\protect\citeauthoryear{Kandula, Orr, and Chaudhuri}{Kandula et~al\mbox{.}}{2019}]%
        {data-induced-predicate}
\bibfield{author}{\bibinfo{person}{Srikanth Kandula}, \bibinfo{person}{Laurel Orr}, {and} \bibinfo{person}{Surajit Chaudhuri}.} \bibinfo{year}{2019}\natexlab{}.
\newblock \showarticletitle{Pushing Data-Induced Predicates through Joins in Big-Data Clusters}.
\newblock \bibinfo{journal}{\emph{Proc. VLDB Endow.}} \bibinfo{volume}{13}, \bibinfo{number}{3} (\bibinfo{date}{nov} \bibinfo{year}{2019}), \bibinfo{pages}{252–265}.
\newblock


\bibitem[\protect\citeauthoryear{Karvounarakis, Green, Ives, and Tannen}{Karvounarakis et~al\mbox{.}}{2013}]%
        {karvounarakis2013collaborative}
\bibfield{author}{\bibinfo{person}{Grigoris Karvounarakis}, \bibinfo{person}{Todd~J Green}, \bibinfo{person}{Zachary~G Ives}, {and} \bibinfo{person}{Val Tannen}.} \bibinfo{year}{2013}\natexlab{}.
\newblock \showarticletitle{Collaborative data sharing via update exchange and provenance}.
\newblock \bibinfo{journal}{\emph{ACM Transactions on Database Systems (TODS)}} \bibinfo{volume}{38}, \bibinfo{number}{3} (\bibinfo{year}{2013}), \bibinfo{pages}{1--42}.
\newblock


\bibitem[\protect\citeauthoryear{Lenzerini}{Lenzerini}{2002}]%
        {lenzerini2002data}
\bibfield{author}{\bibinfo{person}{Maurizio Lenzerini}.} \bibinfo{year}{2002}\natexlab{}.
\newblock \showarticletitle{Data integration: A theoretical perspective}. In \bibinfo{booktitle}{\emph{Proceedings of the twenty-first ACM SIGMOD-SIGACT-SIGART symposium on Principles of database systems}}. \bibinfo{pages}{233--246}.
\newblock


\bibitem[\protect\citeauthoryear{Levy and Mumick}{Levy and Mumick}{1997}]%
        {levy1997query}
\bibfield{author}{\bibinfo{person}{Alon~Yitzchak Levy} {and} \bibinfo{person}{Inderpal~Singh Mumick}.} \bibinfo{year}{1997}\natexlab{}.
\newblock \bibinfo{title}{Query optimization by predicate move-around}.
\newblock
\newblock
\newblock
\shownote{US Patent 5,659,725.}


\bibitem[\protect\citeauthoryear{Meliou, Gatterbauer, Moore, and Suciu}{Meliou et~al\mbox{.}}{2010}]%
        {meliou2010complexity}
\bibfield{author}{\bibinfo{person}{Alexandra Meliou}, \bibinfo{person}{Wolfgang Gatterbauer}, \bibinfo{person}{Katherine~F Moore}, {and} \bibinfo{person}{Dan Suciu}.} \bibinfo{year}{2010}\natexlab{}.
\newblock \showarticletitle{The complexity of causality and responsibility for query answers and non-answers}.
\newblock \bibinfo{journal}{\emph{arXiv preprint arXiv:1009.2021}} (\bibinfo{year}{2010}).
\newblock


\bibitem[\protect\citeauthoryear{Niu, Kapoor, Glavic, Gawlick, Liu, Krishnaswamy, and Radhakrishnan}{Niu et~al\mbox{.}}{2017}]%
        {niu2017provenance}
\bibfield{author}{\bibinfo{person}{Xing Niu}, \bibinfo{person}{Raghav Kapoor}, \bibinfo{person}{Boris Glavic}, \bibinfo{person}{Dieter Gawlick}, \bibinfo{person}{Zhen~Hua Liu}, \bibinfo{person}{Vasudha Krishnaswamy}, {and} \bibinfo{person}{Venkatesh Radhakrishnan}.} \bibinfo{year}{2017}\natexlab{}.
\newblock \showarticletitle{Provenance-aware query optimization}. In \bibinfo{booktitle}{\emph{2017 IEEE 33rd International Conference on Data Engineering (ICDE)}}. IEEE, \bibinfo{pages}{473--484}.
\newblock


\bibitem[\protect\citeauthoryear{Olston, Reed, Srivastava, Kumar, and Tomkins}{Olston et~al\mbox{.}}{2008}]%
        {olston2008pig}
\bibfield{author}{\bibinfo{person}{Christopher Olston}, \bibinfo{person}{Benjamin Reed}, \bibinfo{person}{Utkarsh Srivastava}, \bibinfo{person}{Ravi Kumar}, {and} \bibinfo{person}{Andrew Tomkins}.} \bibinfo{year}{2008}\natexlab{}.
\newblock \showarticletitle{Pig latin: a not-so-foreign language for data processing}. In \bibinfo{booktitle}{\emph{Proceedings of the 2008 ACM SIGMOD international conference on Management of data}}. \bibinfo{pages}{1099--1110}.
\newblock


\bibitem[\protect\citeauthoryear{Psallidas and Wu}{Psallidas and Wu}{2018}]%
        {psallidas2018smoke}
\bibfield{author}{\bibinfo{person}{Fotis Psallidas} {and} \bibinfo{person}{Eugene Wu}.} \bibinfo{year}{2018}\natexlab{}.
\newblock \showarticletitle{Smoke: Fine-grained lineage at interactive speed}.
\newblock \bibinfo{journal}{\emph{arXiv preprint arXiv:1801.07237}} (\bibinfo{year}{2018}).
\newblock


\bibitem[\protect\citeauthoryear{Truong, Sun, Lee, and Guo}{Truong et~al\mbox{.}}{2019}]%
        {truong2019gdpr}
\bibfield{author}{\bibinfo{person}{Nguyen~Binh Truong}, \bibinfo{person}{Kai Sun}, \bibinfo{person}{Gyu~Myoung Lee}, {and} \bibinfo{person}{Yike Guo}.} \bibinfo{year}{2019}\natexlab{}.
\newblock \showarticletitle{Gdpr-compliant personal data management: A blockchain-based solution}.
\newblock \bibinfo{journal}{\emph{IEEE Transactions on Information Forensics and Security}}  \bibinfo{volume}{15} (\bibinfo{year}{2019}), \bibinfo{pages}{1746--1761}.
\newblock


\bibitem[\protect\citeauthoryear{Veanes, Grigorenko, De~Halleux, and Tillmann}{Veanes et~al\mbox{.}}{2009}]%
        {veanes2009symbolic}
\bibfield{author}{\bibinfo{person}{Margus Veanes}, \bibinfo{person}{Pavel Grigorenko}, \bibinfo{person}{Peli De~Halleux}, {and} \bibinfo{person}{Nikolai Tillmann}.} \bibinfo{year}{2009}\natexlab{}.
\newblock \showarticletitle{Symbolic query exploration}. In \bibinfo{booktitle}{\emph{Formal Methods and Software Engineering: 11th International Conference on Formal Engineering Methods ICFEM 2009, Rio de Janeiro, Brazil, December 9-12, 2009. Proceedings 11}}. Springer, \bibinfo{pages}{49--68}.
\newblock


\bibitem[\protect\citeauthoryear{Widom}{Widom}{2004}]%
        {widom2004trio}
\bibfield{author}{\bibinfo{person}{Jennifer Widom}.} \bibinfo{year}{2004}\natexlab{}.
\newblock \bibinfo{booktitle}{\emph{Trio: A system for integrated management of data, accuracy, and lineage}}.
\newblock \bibinfo{type}{{T}echnical {R}eport}. \bibinfo{institution}{Stanford Infolab}.
\newblock


\bibitem[\protect\citeauthoryear{Yan and He}{Yan and He}{2020}]%
        {autosuggest}
\bibfield{author}{\bibinfo{person}{Cong Yan} {and} \bibinfo{person}{Yeye He}.} \bibinfo{year}{2020}\natexlab{}.
\newblock \showarticletitle{Auto-Suggest: Learning-to-Recommend Data Preparation Steps Using Data Science Notebooks}. In \bibinfo{booktitle}{\emph{Proceedings of the 2020 ACM SIGMOD International Conference on Management of Data (SIGMOD)}}. \bibinfo{pages}{1539–1554}.
\newblock


\bibitem[\protect\citeauthoryear{Yan, Lin, and He}{Yan et~al\mbox{.}}{2023}]%
        {predicate-pushdown}
\bibfield{author}{\bibinfo{person}{Cong Yan}, \bibinfo{person}{Yin Lin}, {and} \bibinfo{person}{Yeye He}.} \bibinfo{year}{2023}\natexlab{}.
\newblock \showarticletitle{Predicate Pushdown for Data Science Pipelines}.
\newblock \bibinfo{journal}{\emph{Proceedings of the ACM on Management of Data}} \bibinfo{volume}{1}, \bibinfo{number}{2} (\bibinfo{year}{2023}), \bibinfo{pages}{1--28}.
\newblock


\bibitem[\protect\citeauthoryear{Zaharia, Chowdhury, Franklin, Shenker, Stoica, et~al\mbox{.}}{Zaharia et~al\mbox{.}}{2010}]%
        {zaharia2010spark}
\bibfield{author}{\bibinfo{person}{Matei Zaharia}, \bibinfo{person}{Mosharaf Chowdhury}, \bibinfo{person}{Michael~J Franklin}, \bibinfo{person}{Scott Shenker}, \bibinfo{person}{Ion Stoica}, {et~al\mbox{.}}} \bibinfo{year}{2010}\natexlab{}.
\newblock \showarticletitle{Spark: Cluster computing with working sets.}
\newblock \bibinfo{journal}{\emph{HotCloud}} \bibinfo{volume}{10}, \bibinfo{number}{10-10} (\bibinfo{year}{2010}), \bibinfo{pages}{95}.
\newblock


\bibitem[\protect\citeauthoryear{Zheng, Alawini, and Ives}{Zheng et~al\mbox{.}}{2019}]%
        {zheng2019fine}
\bibfield{author}{\bibinfo{person}{Nan Zheng}, \bibinfo{person}{Abdussalam Alawini}, {and} \bibinfo{person}{Zachary~G Ives}.} \bibinfo{year}{2019}\natexlab{}.
\newblock \showarticletitle{Fine-grained provenance for matching \& ETL}. In \bibinfo{booktitle}{\emph{2019 IEEE 35th International Conference on Data Engineering (ICDE)}}. IEEE, \bibinfo{pages}{184--195}.
\newblock


\bibitem[\protect\citeauthoryear{Zhou, Arulraj, Navathe, Harris, and Wu}{Zhou et~al\mbox{.}}{2021}]%
        {zhou2021sia}
\bibfield{author}{\bibinfo{person}{Qi Zhou}, \bibinfo{person}{Joy Arulraj}, \bibinfo{person}{Shamkant Navathe}, \bibinfo{person}{William Harris}, {and} \bibinfo{person}{Jinpeng Wu}.} \bibinfo{year}{2021}\natexlab{}.
\newblock \showarticletitle{Sia: Optimizing queries using learned predicates}. In \bibinfo{booktitle}{\emph{Proceedings of the 2021 International Conference on Management of Data}}. \bibinfo{pages}{2169--2181}.
\newblock


\end{thebibliography}

\end{document}